\documentclass[10pt, a4paper]{article}
\pdfoutput=1

\usepackage{jcappub}
\usepackage{amssymb}
\usepackage{amsmath}
\usepackage{graphicx}
\usepackage{color}
\usepackage{slashed}
\usepackage{braket} 
\usepackage{hyperref} 	
\usepackage{bm}

\usepackage{myterms}
\newcommand{\CS}{\text{\sc cs}}

\newcommand{\CME}{\text{\sc cme}}

\newcommand{\tinyB}{\text{\sc b}}
\newcommand{\tinyL}{\text{\sc l}}

\newcommand{\tinyY}{\text{\sc y}}

\newcommand{\BE}{\text{\sc be}}

\newcommand{\wt}[1]{\widetilde{#1}}
\renewcommand{\kt}{(\kvec,t)}
\renewcommand{\xt}{(\xvec,t)}
\newcommand{\Avec}{{\bm A}}

\newcommand{\Bvec}{{\bm B}}
\newcommand{\khat}{\hat{\kvec}}
\newcommand{\kabs}{|\kvec|}

\newcommand{\kvec}{{\bm k}}
\newcommand{\pvec}{{\bm p}}
\newcommand{\pabs}{|\pvec|}
\newcommand{\xvec}{{\bm x}}

\newcommand{\dA}{\delta \hspace{-0.04cm} A}
\newcommand{\dB}{\delta \hspace{-0.04cm} B}
\newcommand{\dE}{\delta \hspace{-0.01cm} \Ecal}
\newcommand{\dH}{\delta \hspace{-0.01cm} \Hcal}
\newcommand{\dNCS}{\delta \hspace{-0.04cm} N_{\CS}}
\newcommand{\dof}{{\kappa}}

\def\be{\begin{equation}}
\def\ee{\end{equation}}
\def\ba{\begin{eqnarray}}
\def\ea{\end{eqnarray}}
\def\bald{\begin{aligned}}
\def\eald{\end{aligned}}

\title{\LARGE Chiral Charge Erasure via Thermal Fluctuations of Magnetic Helicity}

\author[a]{\Large Andrew J. Long}
\author[b]{\Large Eray Sabancilar}

\affiliation[a]{Kavli Institute for Cosmological Physics, 
University of Chicago, Chicago, Illinois 60637, USA}
\affiliation[b]{Institut de Th\'eorie des Ph\'enom\'enes Physiques, 
Ecole Polytechnique F\'ed\'erale de Lausanne, CH-1015 Lausanne, Switzerland}

\emailAdd{andrewjlong@kicp.uchicago.edu}
\emailAdd{eray.sabancilar@epfl.ch}

\abstract{
We consider a relativistic plasma of fermions coupled to an Abelian gauge field and carrying a chiral charge asymmetry, which might arise in the early Universe through baryogenesis.  
It is known that on large length scales, $\lambda \gtrsim 1/(\alpha \mu_5)$, the chiral anomaly opens an instability toward the erasure of chiral charge and growth of magnetic helicity.  
Here the chemical potential $\mu_{5}$ parametrizes the chiral asymmetry and $\alpha$ is the fine-structure constant.  
We study the process of chiral charge erasure through the thermal fluctuations of magnetic helicity and contrast with the well-studied phenomenon of Chern-Simons number diffusion.  
Through the fluctuation-dissipation theorem we estimate the amplitude and time scale of helicity fluctuations on the length scale $\lambda$, finding $\dH \sim \lambda T$ and $\tau \sim \alpha \lambda^3 T^2$ for a relativistic plasma at temperature $T$.  
We argue that the presence of a chiral asymmetry allows the helicity to grow diffusively for a time $t \sim T^3/(\alpha^5 \mu_5^4)$ until it reaches an equilibrium value $\Hcal \sim \mu_{5} T^2 / \alpha$, and the chiral asymmetry is partially erased.  
If the chiral asymmetry is small, $\mu_5 < T/\alpha$, this avenue for chiral charge erasure is found to be slower than the chiral magnetic effect for which $t \sim T / (\alpha^3 \mu_{5}^2)$.  
This mechanism for chiral charge erasure can be important for the hypercharge sector of the Standard Model as well as extensions including $\U{1}$ gauge interactions, such as asymmetric dark matter models.
}


\begin{document}
\maketitle

\begingroup
\hypersetup{linkcolor=black}
\tableofcontents
\endgroup

\section{Introduction}\label{sec:Introduction}

There exists a broad class of baryogenesis models in which the departure from thermal equilibrium creates an asymmetry in some anomalous global charge, and the baryon asymmetry is created as various perturbative and non-perturbative reactions act to erase this charge and restore chemical equilibrium.  
Among these models are electroweak baryogenesis \cite{Shaposhnikov:1986jp, Cohen:1990py} and leptogenesis \cite{Fukugita:1986hr, Buchmuller:2004nz}.  
Typically the global charge erasure is accomplished by non-perturbative thermal fluctuations of non-Abelian gauge fields via quantum anomalies; {\it e.g.}, the sum of baryon and lepton numbers is erased by electroweak sphalerons \cite{Kuzmin:1985mm}.  
In this work we consider instead the role of Abelian gauge fields, such as the Standard Model hypercharge field, and explore a new avenue for global charge erasure, namely magnetic helicity diffusion.  
The Abelian case is particularly interesting, because the generation of baryon number could be accompanied by the creation of a relic magnetic field \cite{Vachaspati:1994xc, Cornwall:1997ms, Vachaspati:2001nb}, {\it i.e.} magnetogenesis from baryogenesis, whose detection would serve as an indirect probe of the baryogenesis epoch.  

The fundamental relationship between gauge fields and global charge violation is provided by quantum anomalies \cite{Adler:1969gk,Bell:1969ts}.  
The chiral anomaly manifests itself in the level crossing of chiral fermions in the background of gauge fields with non-trivial topology.  For instance, in QED, as the left and right handed fermion energy levels cross zero energy from above and below respectively, the vector current is still conserved whereas the axial one is not. 
As a result of the chiral anomaly, chiral charge ---the difference between the global charge of the right and left handed fermions--- can be created or destroyed via anomalous processes and the change in chiral charge is related to the change in the topology of the gauge fields. 

To illustrate how non-Abelian and Abelian gauge fields participate in global charge erasure, let us consider a familiar example.  
In the Standard Model both baryon and lepton numbers are anomalous \cite{Hooft:1976up}, as expressed by the divergences of the corresponding currents\footnote{We use standard notation:  $N_f=3$ is the number of Standard Model families, $W_{\mu \nu}$ and $Y_{\mu \nu}$ are the $\SU{2}_L$ and $\U{1}_Y$ field strength tensors, $\wt{W}^{\mu \nu} = (1/2) \epsilon^{\mu \nu \rho \sigma} W_{\rho \sigma}$ and $\wt{Y}^{\mu \nu} = (1/2) \epsilon^{\mu \nu \rho \sigma} Y_{\rho \sigma}$ are the dual tensors, and $g$ and $g^{\prime}$ are the coupling constants.  Recall that ${\rm Tr}\bigl[ F_{\mu \nu} \widetilde{F}^{\mu \nu} \bigr] = (1/2) F_{\mu \nu}^{a} \wt{F}^{a \mu \nu}$ for an $\SU{N}$ field strength tensor $F_{\mu \nu}$. }
\begin{align}\label{eq:djB}
	\partial_{\mu} j^{\mu}_{\tinyB} =\partial_{\mu} j^{\mu}_{\tinyL}= N_f \frac{g^2}{16\pi^2} {\rm Tr}\bigl[ W_{\mu \nu} \widetilde{W}^{\mu \nu} \bigr] - N_f \frac{g^{\prime 2}}{32\pi^2} Y_{\mu \nu} \widetilde{Y}^{\mu \nu} 
	\per
\end{align}
Taking the spacetime integral of \eref{eq:djB} leads to the conservation laws for baryon number $Q_{\tinyB}$ and lepton number $Q_{\tinyL}$.  
The change in these quantities over a finite time interval is given by 
\begin{align}\label{eq:Bdot}
	\Delta Q_{\tinyB} =\Delta Q_{\tinyL} = N_f \Delta N_{\CS} - N_f \frac{g^{\prime 2}}{16\pi^2} \Delta \Hcal_{\tinyY} 
	\com
\end{align}
where $N_{\CS}$ is $\SU{2}_L$ Chern-Simons number [see \eref{eq:NCS_def}] and $\Hcal_{\tinyY} = \int \! \ud^3 x \Avec_{\tinyY} \cdot \Bvec_{\tinyY}$ is the helicity of the hypermagnetic field.  
From \eref{eq:Bdot} we see that changes to either $N_{\CS}$ or $\Hcal$ induce the violation of $B+L$ and thereby facilitate the relaxation of a baryon and lepton number charge asymmetry. 
In a relativistic plasma, thermal fluctuations of the non-Abelian $\SU{2}_L$ gauge field cause Chern-Simons number to diffuse, providing one avenue for baryon and lepton number erasure \cite{Arnold:1987mh, Khlebnikov:1988sr, Mottola:1990bz}.  

In the Abelian sector, the chiral magnetic effect \cite{Vilenkin:1980fu} leads to a growth of magnetic helicity and erasure of the corresponding global charge.  
A chiral charge asymmetry is erased when magnetic helicity is created \cite{Joyce:1997uy,Dvornikov:2013bca,Semikoz:2013xkc,Semikoz:2015wsa,Sabancilar:2013raa,Zadeh:2015oqf}.  
The inverse process, chiral charge creation, and the co-evolution of magnetic helicity have also been studied \cite{Giovannini:1997eg,Giovannini:1999wv,Giovannini:1999by,Dvornikov:2011ey,Dvornikov:2012rk,Semikoz:2012ka,Anber:2015yca,Boyarsky:2011uy,Boyarsky:2012ex,Boyarsky:2015faa,Fujita:2016igl}.  

Here we consider a third avenue for charge erasure, which bears similarities to both non-Abelian Chern-Simons number diffusion and the Abelian chiral magnetic effect.  
In particular, we investigate the thermal fluctuations of magnetic helicity $\Hcal$, the Abelian analog of Chern-Simons number.  
The present study builds on our earlier work, \rref{Long:2013tha}, where we studied magnetogenesis from leptogenesis assuming that magnetic helicity diffusion leads to rapid erasure of summed baryon and lepton number.  
Our thermodynamic arguments and dimensional analysis developed from an analogy with Chern-Simons number diffusion via the $\SU{3}_c$ and $\SU{2}_L$ sphalerons, and we revisit those arguments here.  

The main aspects of our paper can be summarized as follows.  
We find that thermal fluctuations of the Abelian gauge field lead to fluctuations in magnetic helicity $\Hcal$, and we estimate its spectrum using the fluctuation-dissipation theorem.  
We argue that $\Delta \Hcal(t)$ grows diffusively for some time, like the non-Abelian Chern-Simons number, but energetic considerations prevent unbounded growth, and eventually the variance reaches a static equilibrium value.  
If the medium carries an asymmetry in some anomalous global charge, which we call chiral charge in what follows, then helicity diffusion leads to chiral charge erasure on large length scales.  

This paper is organized as follows.  
In \sref{sec:nonAbelian}, we remind the reader how thermal fluctuations of non-Abelian gauge fields lead to chiral charge erasure through Chern-Simons number diffusion and the chiral anomaly.  
This discussion motivates a study of thermal fluctuations in the Abelian sector, which we present in \sref{sec:Abelian_Fluct}, with an emphasis on the diffusion of magnetic helicity.  
Our main results appear in \sref{sec:Abelian_Erasure} where we calculate the time scale on which Abelian thermal fluctuations lead to chiral charge erasure.  
We conclude in \sref{sec:conclusion} and contrast our result with the chiral magnetic effect.  

We use the Heaviside-Lorentz units ($\hbar =1$, $c = 1$) and set the Boltzmann constant $k_B = 1$.  
Boldface type denotes 3-vectors, and their indices are raised and lowered with the Kronecker delta $\delta_{ij} = {\rm diag}(1,1,1)$.  
Normal type denotes 4-vectors, and their indices are raised and lowered with the Minkowski metric $\eta_{\mu \nu} = {\rm diag}(1,-1,-1,-1)$.  
The totally anti-symmetric tensors are normalized as $\epsilon^{0123} = -\epsilon_{0123} = +1$ and $\epsilon^{123} = +1$.

\section{Global Charge Erasure by Diffusion of non-Abelian Chern-Simons Number}
\label{sec:nonAbelian}

In an $\SU{N}$ gauge theory, it is well-known that Chern-Simons number diffuses at finite temperature, and when the theory is coupled to fermions, there is an associated violation of (anomalous) global charge conservation.  
This behavior is manifest, for instance, in the $\SU{3}_c$ and $\SU{2}_L$ sectors of the Standard Model through the so-called strong and electroweak sphalerons.  
We review these phenomena in this section in order to draw a contrast with the behavior of Abelian gauge fields in the following sections.  
We provide citations to the original literature when relevant and refer the reader to the review \cite{Rubakov:1996vz} or the textbook \cite{Shifman:2012} for more details.  

\begin{figure}[t]
\begin{center}
\includegraphics[width=0.48\textwidth]{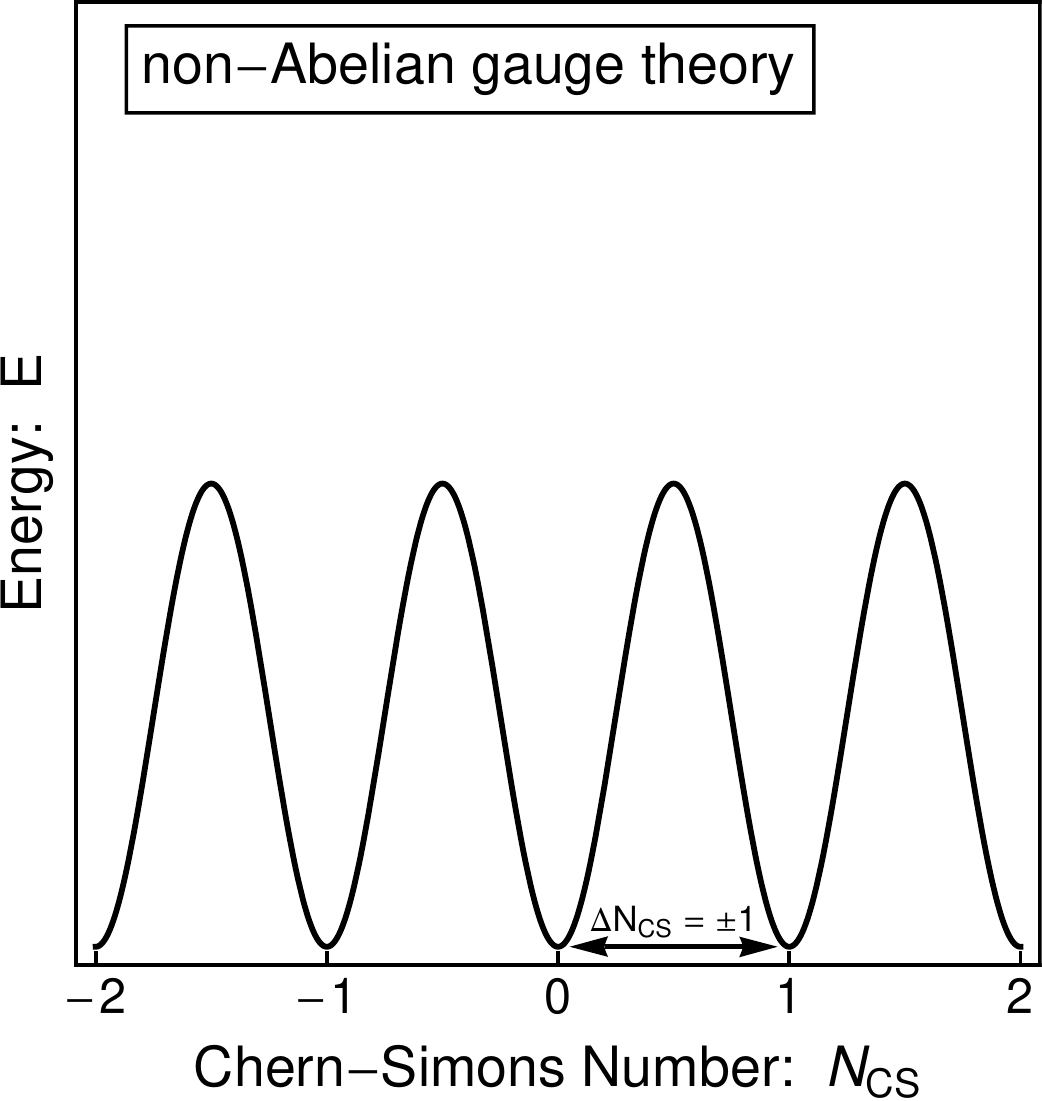} \hfill
\includegraphics[width=0.48\textwidth]{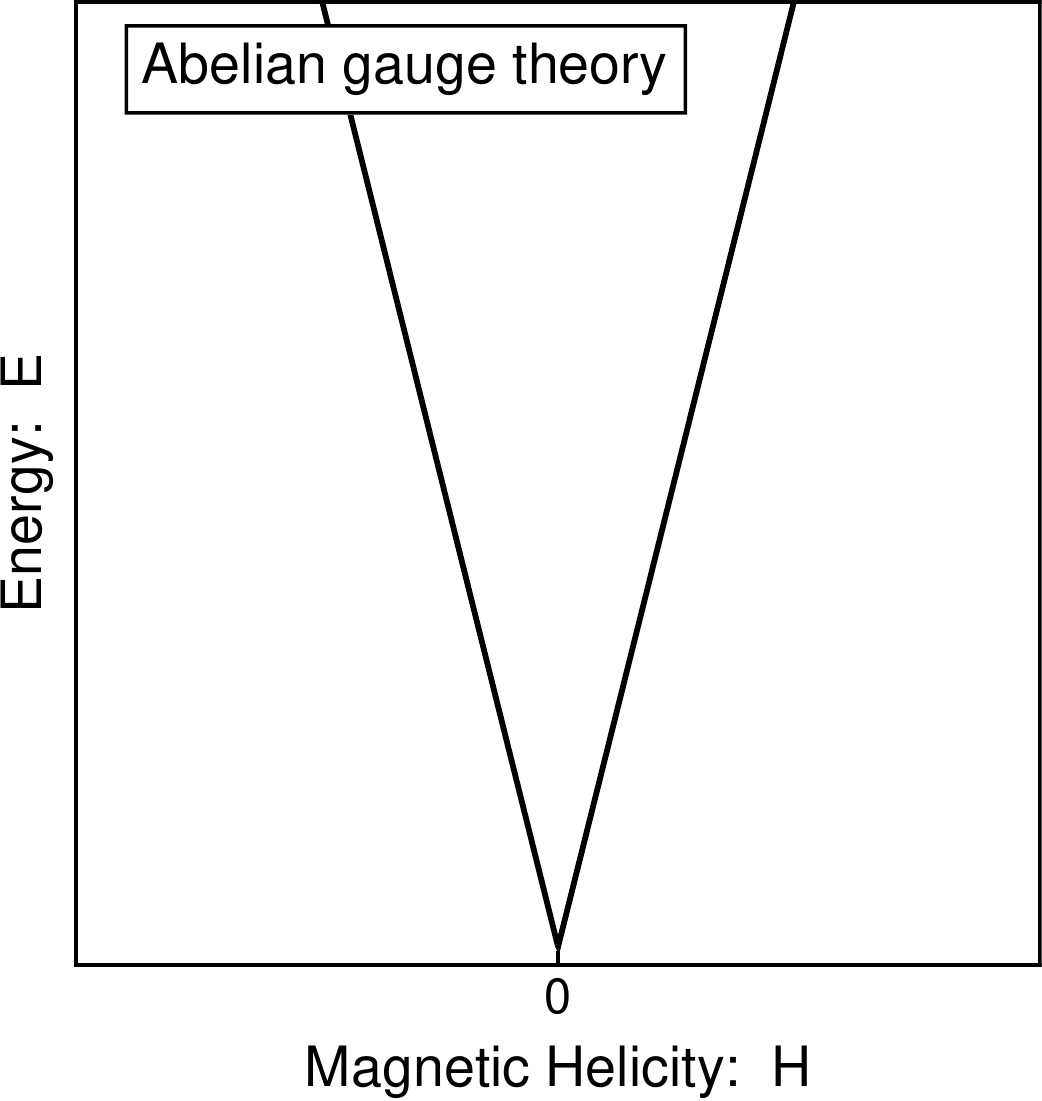} 
\caption{
\label{fig:potentials}
Vacuum structure of the non-Abelian and Abelian gauge theories, represented schematically.  The energy and magnetic helicity have arbitrary units, and the non-Abelian Chern-Simons number changes by one unit between adjacent vacua.  In the Abelian case, the helicity is assumed to saturate the realizability condition, \eref{eq:realizability}, corresponding to a maximally helical magnetic field.  
}
\end{center}
\end{figure}

The vacuum structure of the $\SU{N}$ gauge theory exhibits a family of degenerate vacuum field configurations in 3+1 dimensions \cite{Belavin:1975fg}.  
These configurations are topologically distinct\footnote{The vacuum manifold has a non-trivial third homotopy group $\pi_{3}[{\rm SU}(N)] = \Zbb$.} and identified by their Chern-Simons number, 
\begin{align}\label{eq:NCS_def}
	N_{\CS}(t) 
	= \frac{g^2}{32\pi^2} \int \! \ud^3x \, \epsilon^{0 \alpha \beta \gamma} \bigl( F^a_{\alpha \beta} A^a_{\gamma} - \frac{g}{3} \epsilon_{abc} A^a_\alpha A^b_\beta A^c_\gamma \bigr) \com
\end{align} 
where $A_\mu^a\xt$ is the gauge field, $F^a_{\alpha \beta}\xt$ is the field strength tensor, and $g$ is the gauge coupling.  
Whereas $N_{\CS} \in \Zbb$ is static for the vacuum field configurations, the finite-energy configurations can have varying $N_{\CS} \in \Rbb$, and there exist trajectories through the configuration space that interpolate between the different topological sectors.  
This is illustrated in the left panel of \fref{fig:potentials}.  

\begin{figure}[t]
\begin{center}
\includegraphics[width=0.48\textwidth]{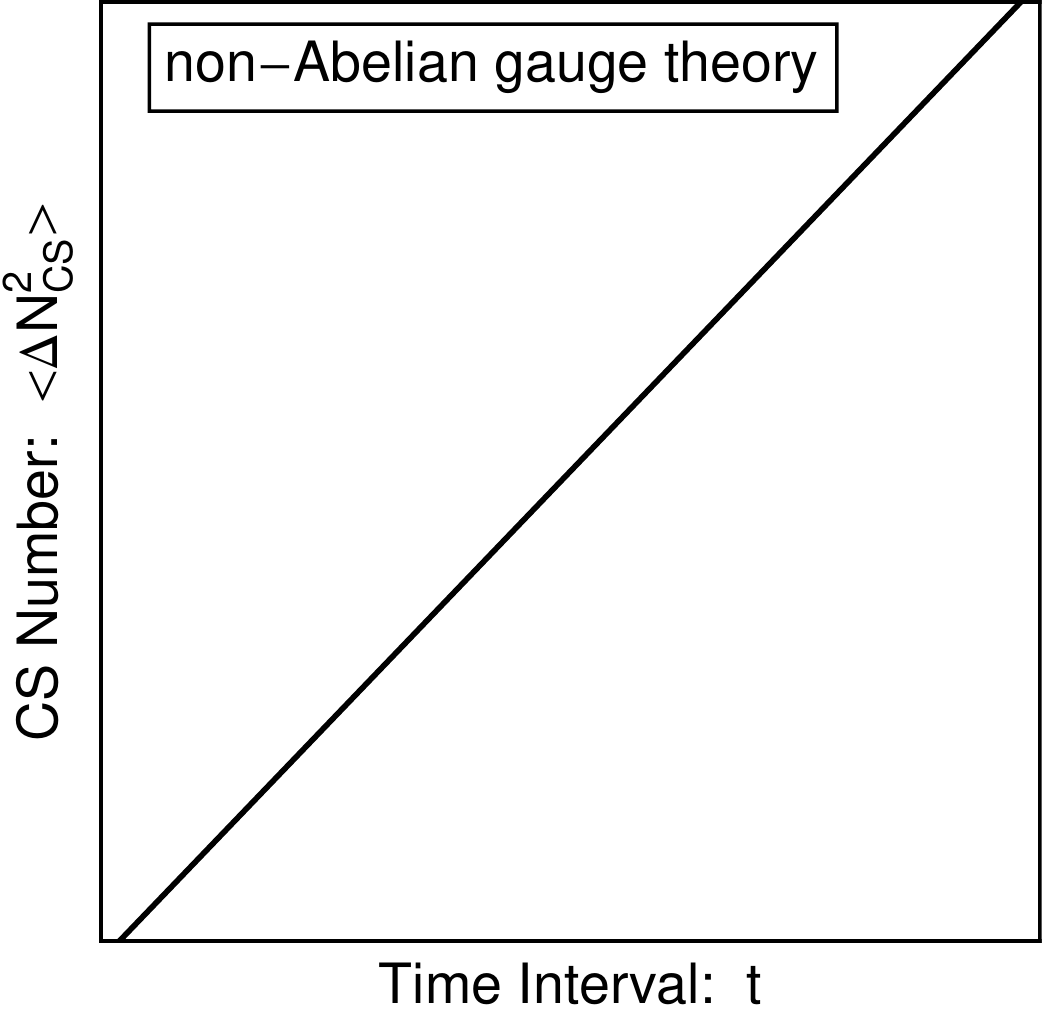} \hfill
\includegraphics[width=0.48\textwidth]{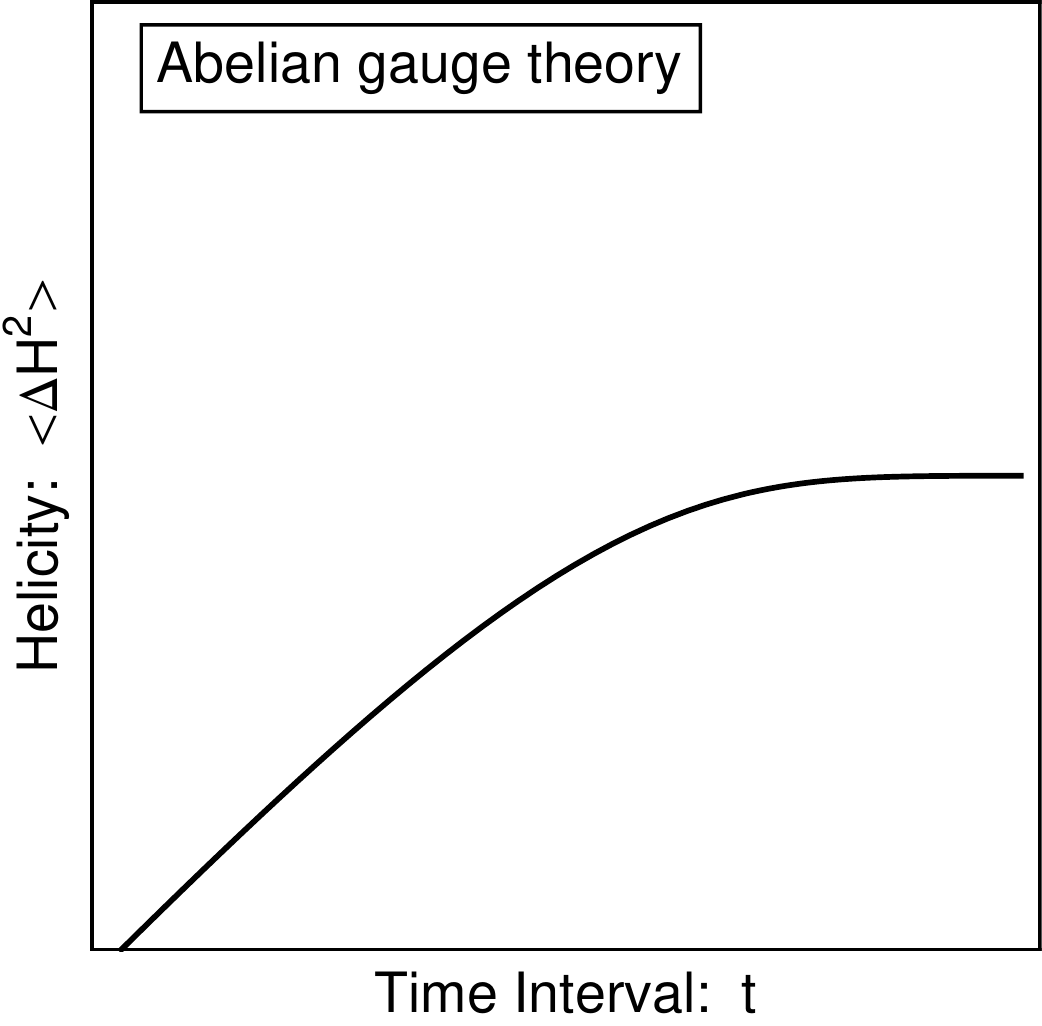} 
\caption{
\label{fig:growth}
Growth of Chern-Simons number and magnetic helicity for the non-Abelian and Abelian gauge theories, represented schematically.  The scales are linear, but the units are arbitrary.  In the absence of fermions, Chern-Simons number can diffuse indefinitely, going as in \eref{eq:NCS_diffusion}, but the growth of magnetic helicity must saturate due to the energy cost of the magnetic field.  
} 
\end{center}
\end{figure}

At finite temperature thermal fluctuations allow $O(1)$ changes in $N_{\CS}$ as the system passes between different topological sectors.  
Since the different sectors are built on degenerate vacua, there is no net energy cost associated with such a transition.  
Consequently $\Delta N_{\CS}(t) = N_{\CS}(t) - N_{\CS}(0)$ performs a random walk in this configuration space.  
That is to say, on time scales $t$ much larger than the rate of fluctuations, we have the diffusive behavior \cite{Khlebnikov:1988sr, Mottola:1990bz} 
\begin{align}\label{eq:NCS_diffusion} 
	\langle \Delta N_{\CS}(t)^2 \rangle = 2 \Gamma V t
\end{align}
with $\Gamma V$ the diffusion coefficient for a system of volume $V$.  
Angled brackets denote thermal averaging.  
The diffusive growth of Chern-Simons number is illustrated in the left panel of \fref{fig:growth}.  

The diffusion coefficient $\Gamma V$ is estimated as follows.  
When crossing between different topological sectors, the system passes close to a saddle point configuration of the free energy\footnote{In the Yang-Mills-Higgs theory with spontaneous symmetry breaking by a scalar field in the fundamental representation of $\SU{N}$, there exists a saddle point configuration of the energy functional known as the sphaleron \cite{Klinkhamer:1984di}.  If the theory is not Higgsed, the energy functional of $\SU{N}$ Yang-Mills does not have a saddle point \cite{Manton:1983nd}, but instead the energy can be lowered monotonically along a particular trajectory \cite{Jackiw:1978zi}.  Nevertheless at finite temperature the free energy acquires a saddle point, which can be viewed as a tradeoff between minimizing the energy and maximizing the entropy \cite{Arnold:1996dy}.  This is the relevant description for $\SU{3}_{c}$ of the SM and $\SU{2}_{L}$ at energies well-above the symmetry breaking scale.  } \cite{McLerran:1990de}.  
As explained in \rref{Arnold:1996dy}, the saddle point configuration can be understood as a non-perturbatively large\footnote{Perturbative fluctuations satisfy $A \sim 1 / \lambda$, but here $A \sim 1 / (\sqrt{\alpha} \lambda)$.} thermal fluctuation with typical spatial extent, temporal duration, and field amplitude given by
\begin{align}\label{eq:ASY_estimate}
	\lambda \sim \frac{1}{\alpha T}
	\qquad , \qquad
	\tau \sim \frac{1}{\alpha^2 T}
	\quad ,  \quad \text{and} \qquad
	\dA \sim \sqrt{\alpha} T \com
\end{align}
where $T$ is the temperature and $\alpha = g^2/4\pi$ is the fine-structure constant for $\SU{N}$.  
These estimates follow immediately from the requirement $\dNCS \sim \alpha \lambda^3 \, \dA \, \dB \sim O(1)$, energetic considerations $\dB^2 \lambda^3 \sim T$, and the fluctuation-dissipation theorem.  
We will utilize similar arguments in \sref{sec:Abelian_Fluct} when discussing the Abelian field fluctuations.  
The Chern-Simons number diffusion coefficient is estimated as $\Gamma V \sim (\dNCS)^2 V / (\lambda^3 \tau)$ leading to \cite{Arnold:1996dy}
\begin{align}\label{eq:Gamma}
	\Gamma \sim \alpha^5 T^4
	\per
\end{align}

If the $\SU{N}$ gauge field is coupled to fermions, some global symmetries will be violated by non-perturbative quantum effects \cite{Adler:1969gk,Bell:1969ts,Hooft:1976up}.  
Typically one talks about symmetries that are conserved at the classical level and refers to the quantum violation as a chiral anomaly, but the symmetry could just as well be violated classically too.  
In the Standard Model, we can identify a fermion-number symmetry for each of the SM fermion species.  
The corresponding Noether currents are violated by the chiral anomalies, 
\begin{align}\label{eq:SM_anomalies}
	\frac{1}{2} \partial_{\mu} j^{\mu}_{Q_L} 
	\ , \ 
	- \partial_{\mu} j^{\mu}_{u_R} 
	\ , \ 
	- \partial_{\mu} j^{\mu}_{d_R} 
	& \supset N_f \frac{g_s^2}{16 \pi^2} \, {\rm Tr}\bigl[ G_{\mu \nu} \, \widetilde{G}^{\mu \nu} \bigr] \com \nn
	\frac{1}{3} \partial_{\mu} j^{\mu}_{Q_L} 
	\ , \ 
	\partial_{\mu} j^{\mu}_{L_L} 
	& \supset N_f \frac{g^2}{16 \pi^2} \, {\rm Tr}\bigl[ W_{\mu \nu} \, \widetilde{W}^{\mu \nu} \bigr] 
	 \com
\end{align}
and also by the Yukawa interactions (not shown).  
The factors of $2$ and $3$ are related to the internal degrees of freedom, weak isospin and color, respectively, and the sign depends on the chirality of the field; see the appendix of \rref{Long:2013tha} for elaboration on the notation.  
When integrated over spacetime volume, the terms on the right side are just the change in Chern-Simons number, $N_f \Delta N_{\CS,c}$ and $N_f \Delta N_{\CS,w}$.  
One can explicitly verify that the $\SU{3}_{c}$ interactions respect the conservation of hypercharge and baryon number, but lead to the anomalous violation of axial baryon number: $\Delta Q_{B5} = -(4/3) N_{f} \Delta N_{\CS,c}$ with $Q_{B5} = \int \! \ud^3x \, \frac{1}{3} \bigl( j_{u_R}^0 + j_{d_R}^0 - j_{Q_L}^0 \bigr)$.  
Similarly the $\SU{2}_{L}$ interactions respect hypercharge conservation, but lead to the violation of\footnote{Other violated charges can be constructed as a linear superposition of $(3B+L)_L$ with any conserved charges.  Violation of left-chiral baryon-plus-lepton number $Q_{(B+L)_L} = \int \! \ud^3x \, \bigl( \frac{1}{3} j_{Q_L}^0 + j_{L_L}^0 \bigr)$ is commonly discussed.  However, it is the combination $(3B+L)_L$, not $(B+L)_L$, that is relaxed to zero by the $\SU{2}_{L}$ anomaly in thermal equilibrium.  } $Q_{(3B+L)_L} = \int \! \ud^3x \, \bigl( j_{Q_L}^0 + j_{L_L}^0 \bigr)$  as $\Delta Q_{(3B+L)_L} = 4 N_{f} \Delta N_{\CS,w}$.  
We will refer to global charges that suffer from a chiral anomaly by the generic term of ``chiral charge'' in what follows.  

The diffusion of the Chern-Simons number provides an avenue for the erasure of chiral charge in a plasma.  
Evolution of the average chiral charge density $n_5 = \langle Q_{5} \rangle / V$ is described with a kinetic equation of the form \cite{Arnold:1987mh, Khlebnikov:1988sr, Mottola:1990bz} 
\ba 
 	\frac{\partial n_5}{\partial t} = - a \, \frac{\Gamma}{T^3} n_5 \com
\ea 
with $\Gamma$ the diffusion coefficient per unit volume from \eref{eq:NCS_diffusion}, and $a$ a dimensionless coefficient.  
Chiral charge erasure via Chern-Simons number diffusion is accomplished on a time scale
\begin{align}\label{eq:tdiff_nonAbel}
	t_{\rm diff} \sim \frac{T^3}{\Gamma} \sim \frac{1}{\alpha^5 T} \com
\end{align}
where we have used $\Gamma \sim \alpha^5 T^4$ from \eref{eq:Gamma}.

\section{Thermal Fluctuations of Abelian Gauge Fields}
\label{sec:Abelian_Fluct}

In \sref{sec:nonAbelian} we reviewed how thermal fluctuations of non-Abelian gauge fields induce chiral charge erasure via Chern-Simons number diffusion.  
We now investigate thermal fluctuation of Abelian gauge fields with an emphasis on magnetic helicity, which is the Abelian analog of Chern-Simons number.
We study the implication for charge erasure in \sref{sec:Abelian_Erasure}.

For an Abelian gauge theory, magnetic helicity is defined as
\begin{align}\label{eq:H_def}
	\Hcal(t)
	= \frac{1}{2}  \int \! \ud^3x \, \epsilon^{0 \alpha \beta \gamma} \bigl( F_{\alpha \beta} A_{\gamma} \bigr) 
	= \int \! \ud^3x \, \Avec\xt \cdot \Bvec\xt 
	\per
\end{align}
The Abelian analog of the non-Abelian Chern-Simons number, \eref{eq:NCS_def}, is related to magnetic helicity as $N_\CS  = (\alpha/ 4\pi) \Hcal $.
As with the Chern-Simons number, $\Hcal$ is gauge invariant if the magnetic field, $\Bvec = {\bm \nabla} \times \Avec$, has no component normal to the boundary of integration.  

We are interested in the spectrum of thermal helicity fluctuations,
\begin{align}\label{eq:heli_fluct}
	\langle \Delta \Hcal(t)^2 \rangle = \langle \bigl( \Hcal(t) - \Hcal(0) \bigr)^2 \rangle 
	\per
\end{align}
Before calculating this quantity, let us draw a contrast between the fluctuations of Abelian helicity and non-Abelian Chern-Simons number.  
Specifically, we saw in \eref{eq:NCS_diffusion} that Chern-Simons number diffuses, $\langle \Delta N_{\CS}^2 \rangle \propto t$, and we might anticipate a similar asymptotic behavior for magnetic helicity, $\langle \Delta \Hcal^2 \rangle \propto t$.  
However, the diffusion of Chern-Simons number is a direct consequence of the unique periodic vacuum structure of non-Abelian $\SU{N}$ gauge theories.  
As we can see in \fref{fig:potentials}, the growth of $N_{\CS}$ does not come with any energy cost.  
The electric and magnetic fields vanish for the vacuum configurations, but $N_{\CS}$ may be nonzero due to the cubic term in \eref{eq:NCS_def}, which is absent in the definition of magnetic helicity in \eref{eq:H_def}.  
On the other hand, in the Abelian sector only finite-energy configurations, $\Bvec \neq 0$, may have nonzero $\Hcal$.  
Clearly the late time behavior cannot be diffusive, $\Delta \Hcal^2 \propto t$, as this would require infinite energy.  
Instead we expect a diffusive behavior at early times and saturation to some temperature-limited asymptotic value at late times.  
This behavior is illustrated schematically in the right panel of \fref{fig:growth}.

Since the relationship between energy and helicity will be important for our calculations, let us take a moment to develop this connection.  
Moving to Fourier space and decomposing onto the circular polarization basis, $B^{\pm}\kt$, we write the magnetic helicity and magnetic energy as \cite{Jackson:1999}
\begin{align}
	\Hcal(t) & = \int \! \ud^3x \, \Avec \cdot \Bvec = \int \! \frac{\ud k}{k} \Hcal_{k}(t)
	\qquad \text{with} \qquad \Hcal_{k}(t) = \frac{1}{k} \int \! \frac{k^2 \ud \Omega_k}{(2\pi)^3} \Bigl( |B^+|^2 - |B^-|^2 \Bigr) 
	\label{eq:H_def_2} \\
	\Ecal(t) & = \frac{1}{2} \int \! \ud^3 x \, |\Bvec|^2 = \int \! \frac{\ud k}{k} \Ecal_{k}(t)
	\qquad \text{with} \qquad \Ecal_{k}(t) = \frac{1}{2} \int \! \frac{k^2 \ud \Omega_k}{(2\pi)^3} \Bigl( |B^+|^2 + |B^-|^2 \Bigr) 
	\label{eq:E_def}
	\per
\end{align}
The so-called realizability condition follows immediately, \cite{FLM:382055}
\begin{align}\label{eq:realizability}
	|\Hcal_{k}(t)| \leq \frac{2}{k} \Ecal_{k}(t) \per
\end{align}
For a given helicity the minimum energy configuration, which saturates the inequality, is said to be maximally helical, {\it i.e.} built up from only one circularization mode or the other.  
The energy cost associated with a maximally helical field is illustrated schematically in the right panel of \fref{fig:potentials}.  

Following the same approach as in \erefs{eq:ASY_estimate}{eq:Gamma} for the non-Abelian case, we will soon estimate the spectrum of thermal fluctuations leading to helicity diffusion.  
One would prefer to evaluate the diffusion coefficient directly by calculating the four-point correlation function in \eref{eq:heli_fluct} with perturbative techniques.  
However, for the non-Abelian case it is well-known that perturbation theory fails to capture Chern-Simons number diffusion \cite{Moore:2000ara,Moore:2010jd}.  
We have already seen in \eref{eq:ASY_estimate} that the thermal fluctuations responsible for Chern-Simons number diffusion are non-perturbatively large, and in \eref{eq:equipartition} we will find the same result for Abelian thermal fluctuations.  
Thus we eschew a direct calculation of the diffusion coefficient, but instead we are guided by the earlier work on Chern-Simons number fluctuations \cite{Arnold:1996dy}.  
We expect that the following estimates give the correct parametric behavior, but the dimensionless coefficients are undetermined.  

Now we estimate the amplitude and rate of thermal fluctuations on a length scale $\lambda = 2\pi / k$.  
The equipartition theorem relates the amplitude of energy fluctuations to temperature, $\dE \sim T$.  
Writing $\dE \sim \lambda^3 (\dB)^2$ as in \eref{eq:E_def} gives the magnetic field and vector potential amplitudes, 
\begin{align}\label{eq:equipartition}
	\dB \sim \sqrt{\frac{T}{\lambda^3}}
	\qquad \text{and} \qquad
	\dA \sim \lambda \dB \sim \sqrt{\frac{T}{\lambda}} \per
\end{align}
Writing $\dH \sim \lambda^3 \, \dA \, \dB$ as in \eref{eq:H_def_2}, we estimate the amplitude of helicity fluctuations as
\begin{align}\label{eq:dH_est}
	\dH \sim \lambda T \per
\end{align}
Note that the realizability condition, \eref{eq:realizability}, is saturated since $\dH \sim \lambda \dE$, and both circular polarization modes are equally represented.  
We are generally interested in length scales much larger than the inter-particle spacing, typically $O(1/T)$ for a relativistic plasma, and thus $\dH \gg 1$ for the scales of interest.  
This should not be surprising; we had the same result in the non-Abelian case where $\dNCS \sim 1$ implies $\dH \sim \dNCS / \alpha \gg 1$ for $\alpha \ll 1$.  
On such large scales, the field fluctuations are non-perturbatively large, $\dA \gg 1 / \lambda$.

\begin{figure}[t]
\begin{center}
\includegraphics[width=0.65\textwidth]{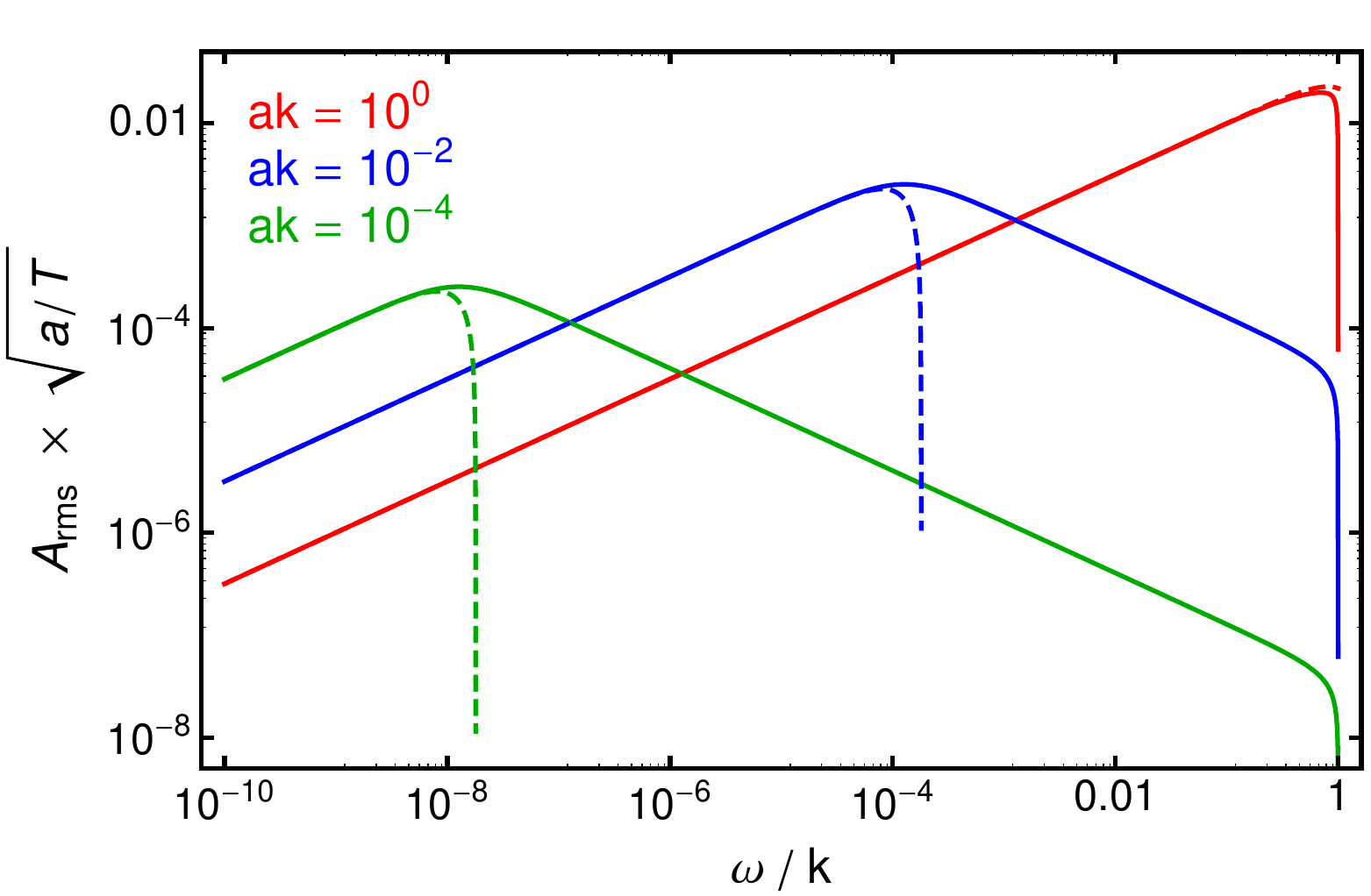} \hfill
\caption{
\label{fig:Arms_spec}
The amplitude of gauge field fluctuations $A_{\rm rms}$ as a function of frequency $\omega$ divided by wavenumber $k = |\kvec|$.  The amplitude is normalized by the ratio of Debye length $a$ and temperature $T$.  We calculate $A_{\rm rms}$ in \eref{eq:Arms_appendix} using the fluctuation-dissipation theorem (solid), and it matches well the approximation in \eref{eq:Arms_approx} (dashed).  
}
\end{center}
\end{figure}

The corresponding fluctuation time scale is obtained from the fluctuation-dissipation theorem\footnote{In linear response theory, the fluctuation-dissipation theorem relates the spectrum of a fluctuating variable to some dissipative property of the system (see e.g., chapter 1 of \cite{Sitenko:1967}). Thus, thermal fluctuations of the gauge field are calculated from the imaginary part of the thermal propagator. We give a quick proof of the fluctuation-dissipation theorem in \aref{app:FDT} and describe the thermal propagators in \aref{app:Rel_Plasma}.
}.  
In \aref{app:Rel_Plasma}, we have calculated the spectrum of thermal fluctuations on the time scale $\tau = 2\pi / \omega$.  
The result is illustrated in \fref{fig:Arms_spec} where we plot the root-mean-square (rms) field amplitude.  
The analytic result is well-approximated to the left of the peak by \eref{eq:Arms_highT_app},
\begin{align}\label{eq:Arms_approx}
	A_{\rm rms}(k,\omega) \sim 
	\sqrt{\frac{T}{a}} \left( 
	\frac{1}{4\pi k} \sqrt{ \frac{1}{2 \pi a} } \omega^{1/2} 
	- \frac{\pi}{128 a^4 k^7} \sqrt{ \frac{1}{2 \pi a} } \omega^{5/2} 
	+ \cdots 
	\right) \com
\end{align}
up to terms that are higher order in $\omega/k \ll 1$ and $a k \ll 1$ and $\omega / T \ll 1$.  
The Debye length evaluates to $a = \sqrt{6/11} \times 1/(g^{\prime} T)$ for the hypercharge plasma of the Standard Model \cite{Arnold:2000dr}.  
The spectrum peaks at a frequency $\omega \sim a^2 k^3$, corresponding to a time scale 
\begin{align}\label{eq:tau_est}
	\tau \sim \frac{\lambda^3}{a^2} 
	\com
\end{align}
and the field amplitude $\dA$ is given by our earlier estimate, \eref{eq:equipartition}.  
One can perform a similar analysis for non-Abelian thermal fluctuations \cite{Arnold:1996dy}.  
In fact, evaluating \erefs{eq:equipartition}{eq:tau_est} with $\lambda \sim 1 / (\alpha T)$ and $a \sim 1/(\sqrt{\alpha} T)$ leads to the spectrum of non-Abelian gauge field fluctuations in \eref{eq:ASY_estimate} and the diffusion coefficient in \eref{eq:Gamma}.  

As we discussed at the beginning of this section, we expect the helicity to grow diffusively on short time scales.  
Using the results above, \erefs{eq:dH_est}{eq:tau_est}, we can estimate
\begin{align}\label{eq:heli_diff_est}
	\langle \Delta \Hcal(t)^2 \rangle 
	\sim (\dH)^2 \frac{t}{\tau} \frac{V}{\lambda^3} 
	\sim 2 \Gamma V t
	\qquad \text{with} \qquad
	\Gamma \sim \frac{a^2 T^2}{\lambda^4 }
	\per
\end{align}
Diffusion will persist until the helicity grows to its equilibrium value.  
In the next section we will see that the presence of a chiral asymmetry induces a large equilibrium value for the magnetic helicity, and the approach to equilibrium lasts for a macroscopically large time.  
If there is no chiral asymmetry, the equilibrium value is comparable to $\dH \sim \lambda T$ as we estimated in \eref{eq:dH_est}, and the diffusive period is brief.

\section{Chiral Charge Erasure by Abelian Helicity Fluctuations}
\label{sec:Abelian_Erasure}

Having estimated the spectrum of magnetic helicity fluctuations in the previous section, we now study the associated chiral charge erasure.  

\subsection{Instability of Chiral Plasma}\label{sec:Energetic}

In this section we demonstrate that a plasma carrying an asymmetry in an anomalous chiral charge is unstable toward the growth of a helical magnetic field.  
We argue from an energetic perspective by calculating the free energy.  
This argument appears in earlier work, {\it e.g.} \rref{Joyce:1997uy}, and our presentation extends and clarifies the argument.  

We consider a QED plasma for ease of discussion and generalize to the hypercharge sector of the Standard Model at the end of this section.  
To set the notation we present the QED Lagrangian:  
\begin{align}\label{eq:L_QED}
	\Lcal & = \overline{\Psi}( i \slashed{D} - m ) \Psi - \frac{1}{4} F_{\mu \nu} F^{\mu \nu} 
\end{align}
with $\Psi\xt$ a Dirac spinor, $A_{\mu}\xt$ a gauge potential, $\slashed{D} \Psi = \gamma^{\mu} ( \partial_{\mu} + i e A_{\mu}) \Psi$, and $F_{\mu \nu} = \partial_{\mu} A_{\nu} - \partial_{\nu} A_{\mu}$.  
The vector, axial-vector, and (Abelian) Chern-Simons current densities are 
\begin{align}
	j^{\mu} = \overline{\Psi} \gamma^{\mu} \Psi 
	\quad , \quad
	j_{5}^{\mu} = \overline{\Psi} \gamma^{\mu} \gamma^{5} \Psi 
	\quad , \quad 
	k^{\mu} = 2 \epsilon^{\mu \alpha \beta \gamma} F_{\alpha \beta} A_{\gamma} 
	\per
\end{align}
An inertial observer with 4-velocity $u^{\mu}=(1,0,0,0)$ identifies the corresponding charge operators, 
\begin{align}
	Q(t) \equiv \int \! \ud^3 x \, j^{0}
	\quad , \quad
	Q_{5}(t) \equiv \int \! \ud^3 x \, j_{5}^{0}
	\quad , \quad 
	\Hcal(t) \equiv \frac{1}{4} \int \! \ud^3 x \, k^{0} \per
\end{align}
These operators represent fermion number (electromagnetic charge per unit $e$), axial fermion number (chiral charge), and electromagnetic helicity, respectively.  
The current densities satisfy 
\begin{align}
	& \partial_{\mu} j^{\mu} = 0 \label{eq:dj} \com \\
	& \partial_{\mu} j_{5}^{\mu}  = 2 i m \overline{\Psi} \gamma^{5} \Psi - \frac{\alpha}{2\pi} F_{\mu \nu} \widetilde{F}^{\mu \nu} \label{eq:djA} \com \\
	& \partial_{\mu} k^{\mu}  = 2 F_{\mu \nu} \widetilde{F}^{\mu \nu} \com
\end{align}
where $\alpha = e^2/4\pi$ is the fine structure constant.  
The first term in \eref{eq:djA} arises from the chirality-violating mass in \eref{eq:L_QED}, and the second term is the Adler-Bell-Jackiw axial anomaly \cite{Adler:1969gk,Bell:1969ts}.  

\begin{figure}[t]
\begin{center}
\includegraphics[width=0.48\textwidth]{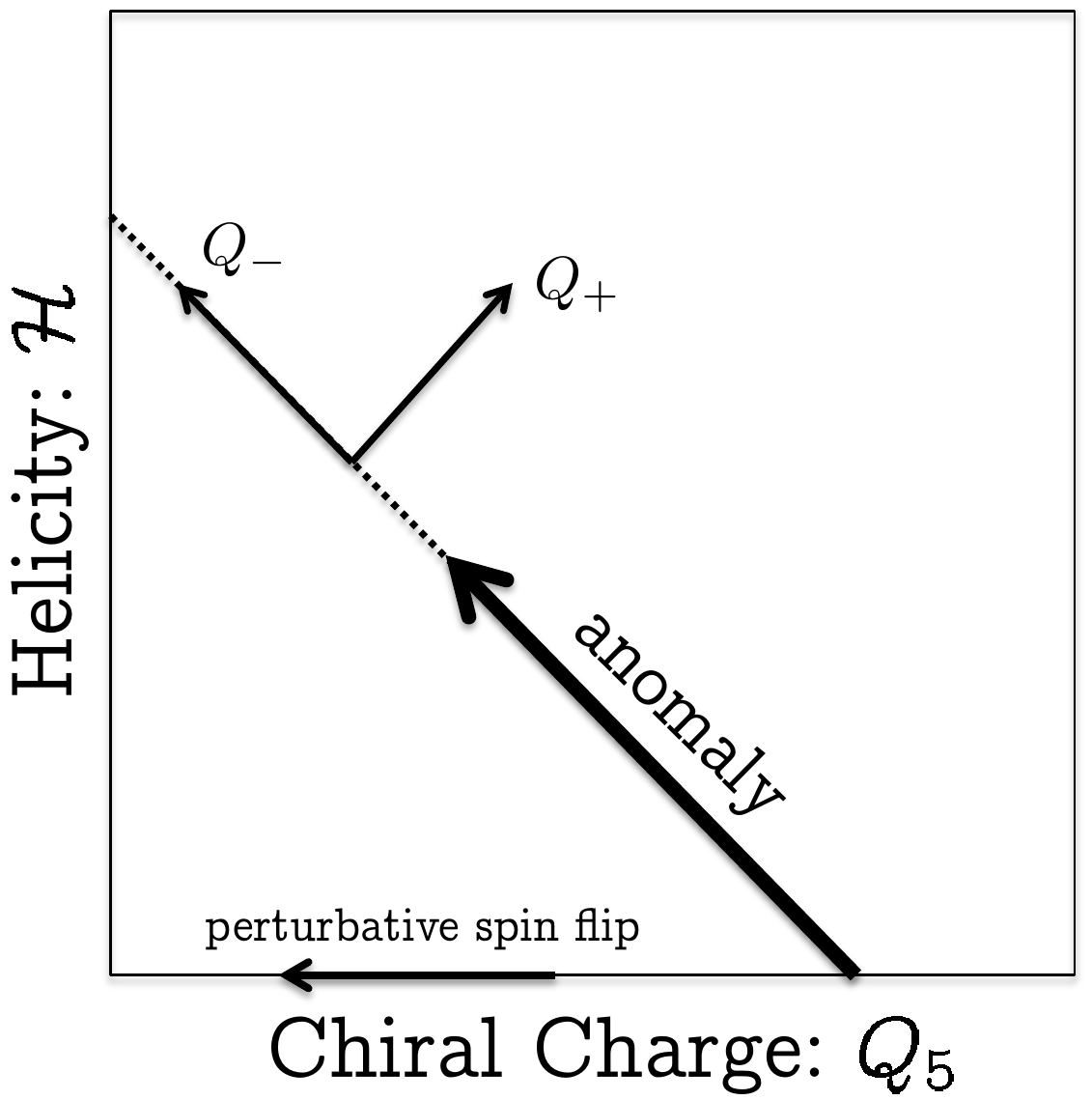} 
\caption{
\label{fig:config_space}
The configuration space parametrized by chiral charge $Q_5$ and helicity $\Hcal$.  The linear combination $Q_{+} = \alpha \Hcal / \pi + Q_{5}$ is conserved, and the orthogonal combination $Q_{-} = \alpha \Hcal / \pi - Q_{5}$ is violated by the anomaly.  We assume that the perturbative spin flip process, which only violates $Q_{5}$, is out of equilibrium.  
}
\end{center}
\end{figure}

Whereas \eref{eq:dj} implies $Q$ is conserved, \eref{eq:djA} reveals that $Q_{5}$ is violated both perturbatively through the electron mass and non-perturbatively through the anomaly.  
Perturbative violation of $Q_{5}$ takes the form of chirality-changing (spin-flip) scattering processes\footnote{The leading order Feynman diagram requires a mass insertion on the internal fermion propagator.}.  
In the early Universe, the rate of these reactions is suppressed by $(m/T)^2$, and at sufficiently high temperature they are out of thermal equilibrium\footnote{The spin-flip interaction remains out of equilibrium for $T \gtrsim 80 \TeV$ \cite{Campbell:1992jd}.  For heavier fermions the interaction comes into equilibrium at a higher temperature.  }.   
We focus on this regime and neglect the electron mass from this point onward. 
Therefore chirality violation is accomplished only through anomaly, and we identify the conserved current $j_5^{\mu} + (\alpha / 4 \pi) k^{\mu}$.  
Thus the system has two conserved charges
\begin{align}\label{eq:consv_charges}
	\Delta Q = 0 
	\qquad \text{and} \qquad
	\frac{\alpha}{\pi} \Delta \Hcal + \Delta Q_5 = 0 \per
\end{align}
It is convenient to introduce
\begin{align}\label{eq:def_Qpm}
	Q_{\pm}(t) \equiv \frac{\alpha}{\pi} \Hcal(t) \pm Q_5(t)
\end{align}
such that $Q_+$ is conserved.  
The configuration space is represented in \fref{fig:config_space}.  
Chirality violation through the anomaly must respect the conservation law in \eref{eq:consv_charges}, but perturbative chirality violation through the spin flip interaction can change $Q_{5}$ directly.  

Consider now a QED plasma with a magnetic field.  
We are interested in how the free energy $F$ varies over the configuration space, parametrized by $Q_5$ and $\Hcal$.  
In the fermion sector there are four degrees of freedom: the two polarization states of the electron ($e_L$, $e_R$) and the positron ($\bar{e}_L$, $\bar{e}_R$).  
In thermal equilibrium, the phase space distribution functions take the Fermi-Dirac form, and we introduce chemical potentials for each degree of freedom.  
Gauge interactions such as $e_L \bar{e}_R \leftrightarrow \gamma \gamma$ occur rapidly and enforce chemical equilibrium, $\mu_{\bar{e}_R} = - \mu_{e_L}$ and $\mu_{\bar{e}_L} = - \mu_{e_R}$.  
The distribution functions can be written as 
\begin{align}
	& f_{e_R} = \Bigl( e^{( E_{\pvec} - \mu_{e_R})/T} + 1 \Bigr)^{-1}
	\quad , \quad
	f_{\bar{e}_L} = \Bigl( e^{( E_{\pvec} + \mu_{e_R})/T} + 1 \Bigr)^{-1} \nn
	& f_{e_L} = \Bigl( e^{( E_{\pvec} - \mu_{e_L})/T} + 1 \Bigr)^{-1}
	\quad , \quad
	f_{\bar{e}_R} = \Bigl( e^{( E_{\pvec} + \mu_{e_L})/T} + 1 \Bigr)^{-1}
	\per
\end{align}
At temperatures $T \gg m$ the energy is approximately $E_\pvec = \sqrt{\pabs^2 + m^2} \approx \pabs$.  
The average fermion number and chiral charge densities are
\begin{align}
	n 
	& = \langle Q \rangle/V = \int \! \! \frac{\ud^3 p}{(2\pi)^3} \Bigl( f_{e_R} - f_{\bar{e}_R} + f_{e_L} - f_{\bar{e}_L} \Bigr) 
	\approx \frac{1}{6} \mu_{Q} T^2 + O(\mu^3) \com
	\label{eq:n_def} \\
	n_5 
	& = \langle Q_5 \rangle/V = \int \! \! \frac{\ud^3 p}{(2\pi)^3} \Bigl( f_{e_R} + f_{\bar{e}_R} - f_{e_L} - f_{\bar{e}_L} \Bigr) 
	\approx \frac{1}{6} \mu_{5} T^2 + O(\mu^3) 
	\label{eq:n5_def} \com
\end{align}
where $\mu_{Q} \equiv \mu_{e_R} + \mu_{e_L}$ and\footnote{More generally the axial chemical potential is calculated as $\mu_{5} = (\mu_{R} - \mu_{L}) / 2$, by summing all right and left handed particles.  The factor of $2$ is canceled to give $\mu_{5} = \mu_{e_R} - \mu_{e_L}$ because of the relation $\mu_{\bar{F}} = - \mu_{F}$ between anti-particle and particle chemical potentials.  Note that $\gamma^5 = P_R - P_L$ without the factor of $2$.  } $\mu_{5} \equiv \mu_{e_R} - \mu_{e_L}$.  
For a neutral plasma $\mu_{Q} = 0$, but we keep the calculation general.  
We have assumed a small chiral asymmetry and dropped terms that are higher order in $\mu/T \ll 1$.  
The average energy density is 
\begin{align}
	\rho = \int \! \! \frac{\ud^3 p}{(2\pi)^3} E_\pvec \Bigl( f_{e_R} + f_{\bar{e}_R} + f_{e_L} + f_{\bar{e}_L} \Bigr) 
	= 4 \times \frac{7}{8} \frac{\pi^2}{30} T^4 + \frac{1}{8} \bigl( \mu_{5}^2 + \mu_{Q}^2 \bigr) T^2 + {\cal O}(\mu^4) \com
\end{align}
where the factor of $4$ counts the degrees of freedom.  
The additional energy density associated to the chiral asymmetry, ${\cal O}(\mu_{5}^2 T^2)$, allows for the generation of magnetic field with strength $B \sim \mu_{5} T$ as we will see below.   
For the relativistic plasma, the pressure is $P = \rho/3$ and the Helmholtz free energy is 
\begin{align}
	F_{\rm fermions} = \bigl( - P + \mu_{e_R} \, n_{e_R} + \mu_{e_L} \, n_{e_L} \bigr) V
	\approx - 4 \times \frac{7}{8} \frac{\pi^2}{90} T^4 V + \frac{1}{24} \bigl( \mu_{5}^2 + \mu_{Q}^2 \bigr) T^2 V + {\cal O}(\mu^4 V) \per
\end{align}
Due to the second term, it is energetically disfavored for the system to maintain the chiral asymmetry if it can instead be relaxed ($\mu_{5} \to 0$).  

If additionally the plasma contains a coherent magnetic field, the corresponding free energy contribution is just its energy $F_{\rm field} = |\Bvec\xt|^2 V/2$.  
To reduce the configuration space to a manageable number of degrees of freedom, we suppose that only one Fourier mode has nonzero amplitude.  
Let the corresponding wavevector be $\kvec = (2 \pi / \lambda) \khat$, and then $F_{\rm field} = B_{\lambda}^2 V/ 2 = \Ecal_{\lambda}$ with $\Ecal_{\lambda}$ the energy carried by the magnetic field.  
In general the magnetic helicity and energy are related by realizability condition, \eref{eq:realizability}, which now takes the form 
\begin{align}\label{eq:realiz_2}
	|\Hcal_{\lambda}| \leq (\lambda/\pi) \mathcal{E}_{\lambda} 
	\per
\end{align}
Thus by assuming that the magnetic field is in a single maximally helical Fourier mode, we reduce the configuration space to four degrees of freedom:  $\khat$, $\lambda$, and $\Hcal_{\lambda}$.  

Combining the free energy of the fermions and the magnetic field gives 
\begin{align}\label{eq:free_energy_1}
	F = \frac{\pi}{\lambda} |\Hcal_{\lambda}| + \frac{1}{24} \mu_{5}^2 T^2 V 
\end{align}
plus terms independent of $\Hcal_{\lambda}$ or $\mu_{5}$.  
If $\Hcal_{\lambda}$ and $\mu_{5}$ could vary independently, the system would evolve toward an equilibrium where $\Hcal_{\lambda}=0$ and $\mu_{5}=0$.  
However, the conservation law in \eref{eq:consv_charges} expresses a constraint requiring changes in $\mu_{5}$ to be compensated by changes in $\Hcal_{\lambda}$.  
To make this explicit, we eliminate $\mu_5$ from \eref{eq:free_energy_1} in favor of the conserved charge 
\begin{align}\label{eq:Q5_to_mu5}
	Q_{+} = \frac{\alpha}{\pi} \Hcal_{\lambda} + \frac{1}{6} \mu_{5} T^2 V \com
\end{align}
given by \erefs{eq:def_Qpm}{eq:n5_def}.  
Doing so gives 
\begin{align}\label{eq:free_energy_2}
	F & = \frac{3}{T^2 V} \frac{\alpha^2}{2\pi^2} \Hcal_{\lambda}^2 + \pi \left( \frac{|\Hcal_{\lambda}|}{\lambda} - \frac{\Hcal_{\lambda}}{\lambda_c} \right) + \frac{3}{2 T^2 V} Q_{+}^2
\end{align}
where we have defined the ``critical'' length scale 
\begin{align}
	\lambda_c \equiv \frac{T^2 V}{3} \frac{\pi^2}{\alpha} \frac{1}{Q_{+}}.  
\end{align}

\begin{figure}[t]
\begin{center}
\includegraphics[width=0.48\textwidth]{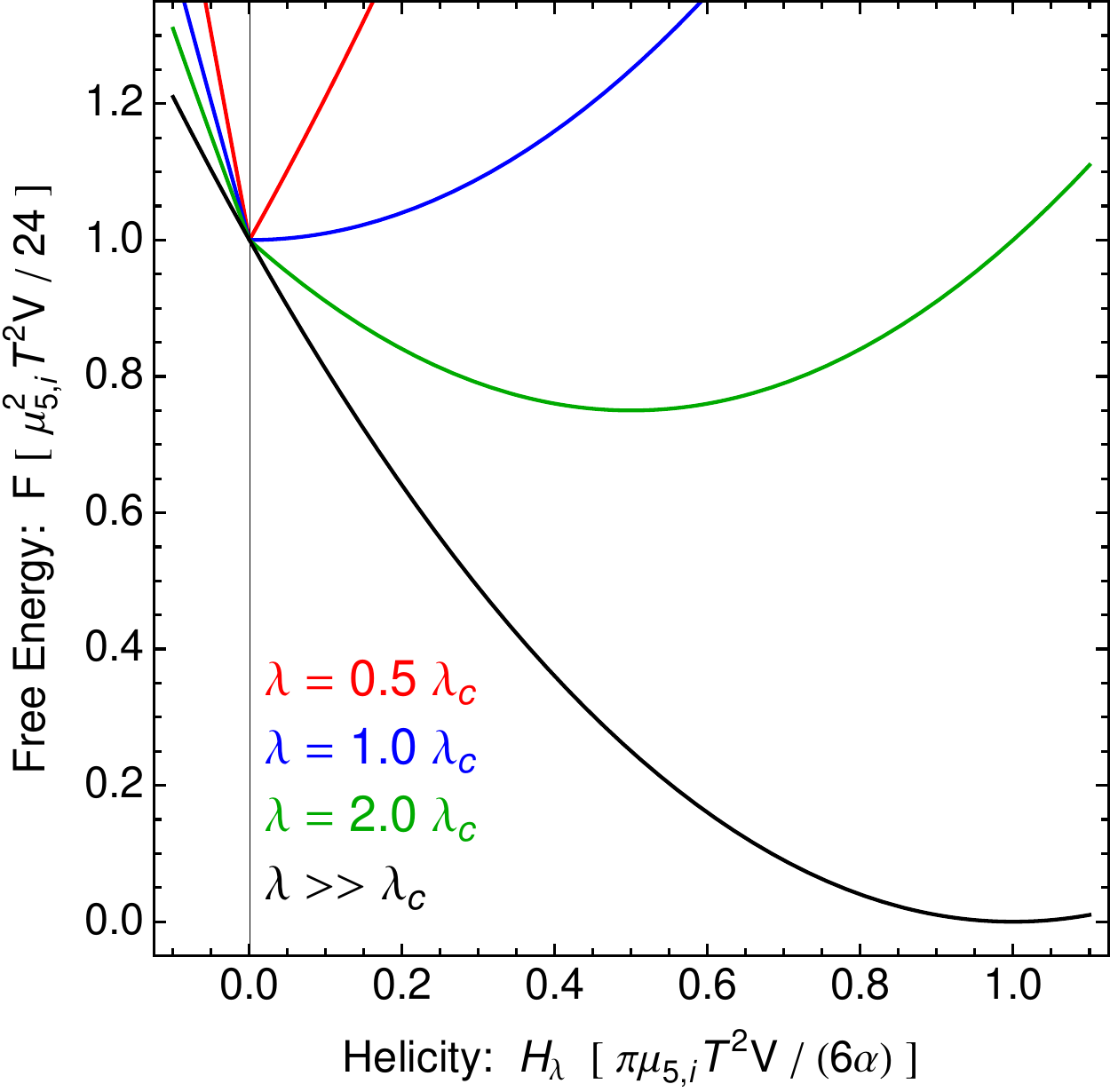} 
\caption{
\label{fig:free_energy}
The free energy given by \eref{eq:free_energy}.  On small length scales, $\lambda < \lambda_{c}$, the free energy is dominated by the energy cost of creating a helical magnetic field, $\Ecal_{\lambda} \approx \pi |\Hcal_{\lambda}|/\lambda$.  On large length scales, $\lambda \gtrsim \lambda_{c}$, the free energy is minimized at $\Hcal_{\lambda} \neq 0$, and it is energetically favorable for the helicity to grow.  
}
\end{center}
\end{figure}

Now suppose that the system is prepared with a chiral asymmetry parametrized by $\mu_{5, i}>0$ and vanishing average magnetic field $\Hcal_{\lambda, i} = 0$.  
Equation~(\ref{eq:Q5_to_mu5}) gives the conserved charge, $Q_{+} = \mu_{5, i} T^2 V / 6$.  
Then the critical length becomes 
\begin{align}\label{eq:lambdac}
	\lambda_{c} = \frac{2\pi^2}{\alpha \mu_{5, i}} 
	\com
\end{align}
and the free energy is 
\begin{align}\label{eq:free_energy}
	F = \frac{3}{T^2 V} \frac{\alpha^2}{2\pi^2} \Hcal_{\lambda}^2 + \pi \left( \frac{|\Hcal_{\lambda}|}{\lambda} - \frac{\Hcal_{\lambda}}{\lambda_c} \right) + \frac{1}{24} \mu_{5, i}^2 T^2 V 
	\per
\end{align}
We minimize the free energy to determine the equilibrium configuration\footnote{Since the system is at finite temperature, it is energetically preferable to minimize the energy $E$ and maximize the entropy $S$ in such a way that the Helmholtz free energy, $F=E-TS=-PV + \mu N$, is minimized as the system approaches thermal equilibrium \cite{FetterWalecka:1971}.  The fermions contribute to both $E$ and $S$.
}.  
This is illustrated graphically in \fref{fig:free_energy}.  
On short length scales, $\lambda < \lambda_c$, the free energy is minimized at $\langle \Hcal_{\lambda}\rangle_{\rm eq}=0$ implying that it is energetically favored to maintain the chiral asymmetry instead of growing the magnetic helicity.  
From the realizability condition, \eref{eq:realiz_2}, there is a large energy cost associated to helicity changes on small length scales.  
On the other hand at large scales, $\lambda > \lambda_c$, the free energy is minimized at 
\begin{align}\label{eq:H_equilib}
	\langle \Hcal_{\lambda} \rangle_{\rm eq} = \frac{\pi}{6 \alpha} \mu_{5, i} T^2 V \left( 1 - \frac{\lambda_c}{\lambda} \right) 
\end{align}
and 
\begin{align}\label{eq:Q5_equilib}
	\langle Q_{5} \rangle_{\rm eq} = \frac{1}{6} \, \mu_{5, i} T^2 V \frac{\lambda_c}{\lambda} 
\end{align}
implying that the system wants to remove some of the chiral charge at the expense of growing the magnetic helicity.  
The corresponding magnetic field strength is estimated as $B_{\lambda} \sim \sqrt{ \Hcal_{\lambda} / (\lambda V)}$ giving
\begin{align}\label{eq:B_equilib}
	\langle B_{\lambda} \rangle_{\rm eq} = \mu_{5,i} T \sqrt{ \frac{1}{12\pi} \left( 1 - \frac{\lambda_c}{\lambda} \right) \frac{\lambda_c}{\lambda} } \per
\end{align}
This completes the instability argument. Note that $\langle \Hcal_{\lambda} \rangle_{\rm eq}$ and $\langle B_{\lambda} \rangle_{\rm eq}$ are not predictions for the spectra, but rather they are the predicted amplitudes of helicity and field strength if the maximally helical magnetic field were carried by a single Fourier mode $k = 2\pi/\lambda$.  
Since all long wavelength modes can be excited $\lambda > \lambda_c$, \erefs{eq:H_equilib}{eq:B_equilib} are more realistically interpreted as upper bounds on the spectra.

The generalization to other Abelian gauge theories is straightforward.  
We assign a chemical potential $\mu_{F}$ to each fermionic particle and a chemical potential $\mu_{\bar{F}}$ to the corresponding anti-particle (CP conjugate).  
Then the chiral chemical potential is generalized to $\mu_{5} = \sum_{x=F,\bar{F}} \chi_x \, q_x^2 \, \mu_x/2$ where $\chi$ is the chirality ($+1$ for right, $-1$ for left) and $q$ is the $\U{1}$ charge.  
Assuming the gauge interactions are in chemical equilibrium, particles in the same gauge multiplet are forced to have the same chemical potential while anti-particles are just the opposite, $\mu_{\bar{F}} = - \mu_{F}$.  
The the chiral chemical potential is calculated by taking one representative particle from each multiplet, \cite{Fukushima:2012vr}
\begin{align}\label{eq:mu5_general}
	\mu_5 = \sum_f ~\dof_{f}~\chi_f~ q_f^2 ~\mu_f 
	\com
\end{align}
and $\dof_{f}$ counts the internal degrees of freedom ({\it e.g.}, color and isospin).  
For QED we have $f = \{ e_R \, , \, e_L \}$ with $\dof_{e_R} = \dof_{e_L} = 1$, $\chi_{e_R} = - \chi_{e_L} = 1$, $q_{e_R} = q_{e_L} = 1$, and thus $\mu_{5} = \mu_{e_R} - \mu_{e_L}$ as in \eref{eq:n5_def}.  

For the hypercharge sector of the SM we replace the QED axial anomaly, \eref{eq:djA}, with the set of chiral anomalies
\begin{align}\label{eq:U1Y_anomalies}
	- \frac{\partial_{\mu} j^{\mu}_{u_R}}{3 (y_{u_R}/2)^2} 
	\ , \ 
	- \frac{\partial_{\mu} j^{\mu}_{d_R}}{3 (y_{d_R}/2)^2} 
	\ , \ 
	- \frac{\partial_{\mu} j^{\mu}_{e_R}}{(y_{e_R}/2)^2}  
	\ , \ 
	\frac{\partial_{\mu} j^{\mu}_{Q_L}}{6 (y_{Q_L}/2)^2}  
	\ , \ 
	\frac{\partial_{\mu} j^{\mu}_{L_L}}{2 (y_{L_L}/2)^2} 
	\supset N_f \frac{g^{\prime 2}}{16 \pi^2} \, Y_{\mu \nu} \, \widetilde{Y}^{\mu \nu} 
	\per
\end{align}
These can be compared with \eref{eq:SM_anomalies} for the non-Abelian sectors.  
The factors of $\kappa_f = 2,3$ and $6$ count the internal degrees of freedom, and the charges under $\U{1}_Y$ are $y_{Q_L} = 1/3$, $y_{u_R} = 4/3$, $y_{d_R} = -2/3$, $y_{L_L} = -1$, and $y_{e_R} = -2$.  
The anomalies in \eref{eq:U1Y_anomalies} respect four conservation laws (among them are hypercharge and baryon-minus-lepton number) and lead to the anomalous violation of hypercharge-weighted chiral charge, $Q_{5} = \int \! \ud^3x \bigl( -y_{Q_L}^2 j_{Q_L}^0 + y_{u_R}^2 j_{u_R}^0 + y_{d_R}^2 j_{d_R}^0 - y_{L_L}^2 j_{L_L}^0 + y_{e_R}^2 j_{e_R}^0 \bigr)$.  
Then \eref{eq:mu5_general} gives the chiral chemical potential to be
\begin{align}\label{eq:muY}
	\mu_5 
	& = N_f \Bigl[ 
	3 \left( \frac{4}{3} \right)^2 \mu_{u_R} + 3 \left( -\frac{2}{3} \right)^2 \mu_{d_R} + \left( -2 \right)^2 \mu_{e_R} - 6 \left( \frac{1}{3} \right)^2 \mu_{Q_L} - 2 \left( -1 \right)^2 \mu_{L_L} 
	\Bigr]  \com \nn
	& = \frac{N_f}{3} \Bigl[ 16 \mu_{u_R} + 4 \mu_{d_R} + 12 \mu_{e_R} - 2 \mu_{Q_L} - 6 \mu_{L_L} \Bigr]
	\per
\end{align}
While we refer to charge associated with $\mu_{5}$ as a chiral charge, it could be more appropriately described as a squared-hypercharge-weighted chiral charge.  

\subsection{Diffusive Approach to Equilibrium}\label{sec:Approach}

We now estimate the time scale $t_{\rm diff}$ required for the system to reach equilibrium via diffusive thermal fluctuations of magnetic helicity.  
We previously studied helicity fluctuations in \sref{sec:Abelian_Fluct}.  
Focusing now on the instability length scale identified in \eref{eq:lambdac}, we have the length scale, time scale, and amplitude of helicity fluctuations from \erefs{eq:dH_est}{eq:tau_est} to be 
\begin{align}\label{eq:deltaHc_est}
	\lambda_c \sim \frac{1}{\alpha \mu_{5}}
	\qquad , \qquad
	\tau_c \sim \frac{1}{\alpha^3 \mu_{5}^3 a^2}
	\quad , \quad \text{and} \qquad
	\dH_c 
	\sim \frac{T}{\alpha \mu_{5}}
	\per
\end{align}
The diffusive growth of magnetic helicity was estimated in \eref{eq:heli_diff_est}, and for the length scale of interest it evaluates to 
\begin{align}\label{eq:Hvar_diff}
	\langle \Delta \Hcal_{\lambda_c}^2 \rangle 
	\sim (\dH_c)^2 \frac{t}{\tau_c} \frac{V}{\lambda_c^3} 
	\sim 2 \Gamma_c V t
	\qquad \text{with} \qquad
	\Gamma_c \sim a^2 T^2 \alpha^4 \mu_{5}^4
	\per
\end{align}
Recalling that $a^2 \sim 1/(\alpha T^2)$, we find the diffusion coefficient to be $\Gamma_c V \sim \alpha^3 \mu_{5}^4 V$.  

We expect that the magnetic helicity will continue to grow until it reaches its equilibrium value.  
When the system reaches thermal equilibrium, the probability distribution for $\Hcal_{\lambda}$ is proportional to the Boltzmann factor, ${\rm exp}[- F/T]$, with the free energy given by \eref{eq:free_energy}.  
For $\lambda > \lambda_c$ this is a Gaussian distribution with variance
\begin{align}\label{eq:Hvar_eq}
	\langle \Hcal_{\lambda}^2 \rangle_{\rm eq} = \frac{T^3 V}{9} \frac{2\pi^2}{\alpha^2} \per
\end{align}
Equating \erefs{eq:Hvar_diff}{eq:Hvar_eq} we estimate the time to reach equilibrium\footnote{These estimates are motivated by an analogy with the one-dimensional random walk in a parabolic confining potential, {\it i.e.} the Ornstein-Uhlenbeck process \cite{GrimmettStirzaker:2001}.  The probability distribution for the particle's location evolves subject to the Smoluchowski equation \cite{2005CoTPh..43.1099S} with a delta-function initial condition.  The variance of the displacement grows linearly until it reaches the equilibrium value after a time $t \sim \langle x^2 \rangle_{\rm eq} / D$ with $D$ the diffusion coefficient. }, 
\begin{align}\label{eq:tdiff_est}
	t_{\rm diff} \sim \frac{\langle \Hcal_{\lambda}^2 \rangle_{\rm eq}}{\Gamma_c V} \sim \frac{T}{a^2 \alpha^6 \mu_{5}^4} 
	\per
\end{align}
This estimate is one of the main results of our paper.  
It reveals that the diffusive approach to equilibrium is very slow if the initial chiral asymmetry is small, going like $1/\mu_{5}^4$.  
This result should be compared with \eref{eq:tdiff_nonAbel} for the case of Chern-Simons number diffusion.  
However, we note that the anomalous charges violated in the non-Abelian and Abelian cases will typically differ.  
In particular, for the case of SM hypercharge sector, the chiral charge given by \eref{eq:muY} is relaxed during the diffusion.  

\section{Summary and Discussion}\label{sec:conclusion}

In this paper we have provided a new perspective on chiral charge erasure via the diffusion of magnetic helicity.  
Our approach and results are summarized as follows.  
We estimated the spectrum of thermal magnetic helicity fluctuations using the fluctuation-dissipation theorem in \sref{sec:Abelian_Fluct}.  
We argued that the magnetic helicity grows diffusively, $\langle \bigl( \Hcal(t) - \Hcal(t^{\prime}) \bigr)^2 \rangle \propto t-t^{\prime}$, until it reaches an equilibrium value determined by energetic limitations.  
In the presence of a chiral asymmetry with chemical potential $\mu_{5}$, the equilibrium value is displaced from zero.  
On large length scales, $\lambda \gtrsim 1/(\alpha \mu_5)$ given by \eref{eq:lambdac}, the equilibrium magnetic helicity and field strength are given by \erefs{eq:H_equilib}{eq:B_equilib} as $\Hcal \sim \mu_{5} T^2 V / \alpha$ and $B \sim \mu_{5} T$.  
The system approaches equilibrium through helicity diffusion on a time scale $t \sim T / (a^2 \alpha^6 \mu_{5}^4)$ given by \eref{eq:tdiff_est}, and $a \sim 1/(\sqrt{\alpha}T)$ is the Debye length.  
Due to the chiral anomaly, helicity diffusion is accompanied by a partial erasure of the initial chiral charge, given by \eref{eq:Q5_equilib}.
On short length scales, $\lambda \lesssim 1/(\alpha \mu_{5})$, it is energetically preferable for the system to retain its chiral charge without the generation of magnetic helicity.  

\begin{table}[t]
\begin{center}
\begin{tabular}{|l|c|c|}
\hline
 & non-Abelian & Abelian \\
\hline 
$\begin{array}{c}\text{Fluctuation length scale relevant for charge erasure ($\lambda_{c}$)} \end{array}$ 
& $\frac{1}{\alpha_{\rm nA} T}$
& $\frac{1}{\alpha_{\rm A} \mu_{5}}$\\
\hline
$\begin{array}{c}\text{Fluctuation time scale relevant for charge erasure ($\tau_{c}$)} \end{array}$ 
& $\frac{1}{\alpha_{\rm nA}^2 T}$
& $\frac{T^2}{\alpha_{\rm A}^{2} \mu_{5}^3}$\\
\hline
$\begin{array}{c}\text{Time scale to accomplish charge erasure ($t_{\rm diff}$)} \end{array}$ 
& $\frac{1}{\alpha_{\rm nA}^5 T}$
& $\frac{T^3}{\alpha_{\rm A}^{5} \mu_{5}^4}$\\
\hline
\end{tabular}
\end{center}
\caption{\label{tab:comparison}
Comparison of non-Abelian $\SU{N}$ and Abelian $\U{1}$ thermal fluctuations leading to diffusion of Chern-Simons number and magnetic helicity at temperature $T$.  The non-Abelian and Abelian gauge couplings enter the corresponding fine-structure constant, given by $\alpha = g^{2}/(4\pi)$, and we have written the Debye length as $a \sim 1 / (\sqrt{\alpha} T)$.  In deriving the Abelian relations we have assumed a small chiral charge asymmetry, $\mu_{5} < T$.  
}
\end{table}
%

Chiral charge erasure by magnetic helicity diffusion is different in many aspects from non-Abelian Chern-Simons number diffusion:  
\begin{enumerate}
	\item  
We compare the relevant length and time scales in \tref{tab:comparison}.  
In the non-Abelian case, $1/(\alpha T)$ is the shortest length scale on which thermal fluctuations satisfy $\dNCS \sim O(1)$.  
In the Abelian case, charge erasure is only energetically favorable on large length scales where the energy cost of creating a magnetic field is smaller than the energy liberated in erasing the chiral asymmetry.  
For $\mu_{5} \ll T$, which is typically the case for asymmetries arising from baryogenesis, charge erasure is much slower in the Abelian case.  
	\item  
In the non-Abelian case, Chern-Simons number can be carried by the vacuum sector, and therefore we do not expect Chern-Simons number diffusion to produce radiation of a non-Abelian magnetic field while the system remains in thermal equilibrium\footnote{However at the electroweak phase transition it has been argued that Chern-Simons number in $\SU{2}_{L}$ can convert into magnetic helicity of $\U{1}_{em}$ \cite{Jackiw:1999bd}.  }.  
In the Abelian gauge theory, the chiral anomaly requires changes in chiral charge to be compensated by changes in magnetic helicity, \eref{eq:consv_charges}, and thus charge erasure is accompanied by helicity generation.  
We expect this magnetic helicity to be associated with the radiation of a coherent magnetic field, because the chiral anomaly is non-perturbative in nature.  
This assumption has been implicit in our calculations.  
If the plasma acquires a helicity without the creation of a coherent magnetic field, perhaps because the helicity is carried by the photons comprising the thermal bath, then the helicity--energy relation in \eref{eq:realiz_2} does not hold, and the energetic argument of \sref{sec:Energetic} breaks down.  
Specifically, if helicity is created without any energy cost, then helicity diffusion will resemble the diffusion of Chern-Simons number.  
	\item  
If the non-Abelian theory is not coupled to fermions, Chern-Simons number diffusion can continue indefinitely since there is no energy cost associated with $O(1)$ changes in $N_{\CS}$ due to the periodic vacuum structure of the theory.  
Coupling the theory to fermions, which is the case for the SM gauge groups anyway, the diffusion of $N_{\CS}$ is accompanied by a diffusion of the associated anomalous global charge.  
Since there is an energy cost to this global charge asymmetry, $\mu^2 T^2 V$, the diffusion cannot continue indefinitely \cite{Khlebnikov:1988sr, Mottola:1990bz}.  
During $N_{\CS}$ diffusion and subsequent erasure of the chiral asymmetry, the energy in fermions simply thermalize and an additional entropy is injected to the system.  
In the Abelian sector the magnetic helicity carried by a coherent magnetic field is also unable to grow indefinitely, but this is the case even if the theory were not coupled to fermions.  
This is because a magnetic field with helicity $\Hcal$ costs an energy $\Ecal \gtrsim \Hcal / \lambda$, as in \eref{eq:realizability}.  
	\item  
The chiral anomalies of the Standard Model, \eref{eq:SM_anomalies}, imply that diffusion of $N_{\CS}$ in the $\SU{3}_{c}$ and $\SU{2}_{L}$ sectors leads to the erasure of axial baryon number and left-chiral baryon-plus-lepton number, respectively.  
Magnetic helicity diffusion in the $\U{1}_Y$ sector tends to erase the squared-hypercharge-weighted chirality, given by \eref{eq:U1Y_anomalies}.  
\end{enumerate}

The diffusive growth of magnetic helicity in an Abelian plasma can be studied more rigorously using lattice gauge theory techniques.  
The quantity of interest is \cite{Mottola:1990bz}
\begin{align}\label{N_cs lattice}
	\Gamma(t) = \frac{1}{2} \lim_{V \to \infty} \int_{-t}^{t} \ud t^{\prime} \int_V \ud^3 x^{\prime} \ \frac{1}{2} \bigl< q(x^{\prime}) q(0) + q(0) q(x^{\prime}) \bigr>
\end{align}
with $q(x) \equiv g^2/(16\pi^2) \, Y_{\mu \nu} \wt{Y}^{\mu \nu}$.  
For non-Abelian fields in the absence of fermions, $\lim_{t \to \infty} \Gamma(t)$ gives the diffusion coefficient appearing in \eref{eq:NCS_diffusion}.  
For Abelian fields we expect that $\Gamma(t) \sim 1/t$ and $\langle \Delta \Hcal^2 \rangle = 2 \Gamma V t$ is static at late times, because helicity fluctuations reach their equilibrium value, as we argued in \sref{sec:Abelian_Fluct}.  
It would be interesting to study the evolution of magnetic helicity in the presence of a chiral asymmetry using lattice techniques.  

The growth of magnetic helicity can also be studied in the formalism of chiral MHD \cite{Boyarsky:2015faa}.  
The combination of a chiral asymmetry $\mu_{5}$ and background magnetic field $\Bvec$ lead to a non-dissipative electric current ${\bf j} = (2\alpha/\pi) \mu_{5} \Bvec$, via the chiral magnetic effect \cite{Vilenkin:1980fu}.  
Combining Maxwell's equations with the non-dissipative current and Ohm's law leads to the induction equation 
\begin{align}\label{eq:induction_eqn}
	\dot\Bvec = \frac{1}{\sigma} \nabla^2 \Bvec + \frac{2\alpha}{\pi} \frac{\mu_{5}}{\sigma} {\bm \nabla} \times \Bvec \per
\end{align}
The conductivity is $\sigma \sim T / \alpha$ for a relativistic plasma \cite{Baym:1997gq}.  
This linear equation is easily solved in Fourier space after decomposing onto the circular polarization basis:  
\begin{align}\label{eq:induction_soln}
	B^{\pm}\kt = B^{\pm}(\kvec,t_i) \, {\rm exp}\Bigl[ - \frac{1}{\sigma} \bigl( k \mp k_c \bigr)^2 (t-t_i) \Bigr] \, {\rm exp}\Bigl[ \frac{1}{\sigma} k_c^2 (t-t_i) \Bigr] \com
\end{align}
where $\bar{\mu}_5$ is the time average of $\mu_{5}$.  
The instability scale, $k_c = \alpha \bar{\mu}_5 / \pi$ or $\lambda_{c} = 2\pi^2 / (\alpha \bar{\mu}_5)$, is the same one we encountered from the purely energetic arguments leading to \eref{eq:lambdac}.  
One of the helicity modes is unstable toward an exponential amplification on large length scales $1/k > 1/k_c$, and the instability develops on a time scale 
\begin{align}
	t_{\CME} \sim \frac{\sigma}{k_c^2} \sim \frac{T}{\alpha^3 \bar{\mu}_{5}^2} \per
\end{align}
Comparing with the diffusion time scale in \eref{eq:tdiff_est}, we see that generally $t_{\CME} / t_{\rm diff} \sim (\alpha \bar{\mu}_{5}/T)^2 < 1$ for $\bar{\mu}_{5} < T/\alpha$, as we have assumed.  
Then equilibrium is always reached through the MHD instability before it is reached by diffusion.
However, it remains to be seen whether the diffusive growth of magnetic helicity can have practical applications in a plasma with large chiral asymmetry (see e.g., Ref.~\cite{Rubakov:1986am, Rubakov:1985nk,Akamatsu:2013pjd}, where chiral instabilities have been studied in the presence of a large chiral asymmetry.).

In comparing helicity diffusion and the chiral magnetic effect, it is worth emphasizing that both avenues for magnetic field generation rely on the chiral anomaly and must satisfy the conservation law $\Delta \Hcal = - (\alpha/\pi) \Delta Q_{5}$.  
Although the full spectrum of the magnetic field will depend on the dynamics of its generation, we see in this equation that the sign of the magnetic helicity is determined by the sign of the initial chiral asymmetry in a model-independent way.  
Since the relationship with the baryon asymmetry is model-dependent, one can hope to use measurements of the relic primordial magnetic field to discriminate between models of baryogenesis, for instance using TeV blazars and the diffuse gamma ray data \cite{Neronov:1900zz,Taylor:2011bn,Tashiro:2013ita,Chen:2014qva}.  

Chiral charge erasure via helicity diffusion may have implications for early Universe cosmology.  
We compare the helicity diffusion time scale, $t_{\rm diff}$ from \eref{eq:tdiff_est}, with the age of the Universe $t_U$.  
During radiation domination, $t_U \sim H^{-1} \sim M_{\rm pl}/T^2$ with $H$ the Hubble parameter and $M_{\rm pl} \sim 10^{18} \GeV$ the Planck mass.  
Charge erasure is accomplished through helicity diffusion at temperature $T$ provided that $t_{\rm diff}<t_U$ or 
\begin{align}\label{eq:tdiff_bound}
	\frac{\mu_{5}}{T} \gtrsim \left( \frac{T}{M_{\rm pl}} \right)^{1/4} \frac{1}{\alpha^{5/4}} 
	\sim 10^{-1} \left( \frac{T}{100 \TeV} \right)^{1/4} \per
\end{align}
We use $\alpha \sim 1/100$ in the estimate.  
We have normalized to $100 \TeV$ since the perturbative spin-flip interaction comes into equilibrium below this scale \cite{Campbell:1992jd} and tends to erase the chiral asymmetry $\mu_5$ without an associated generation of magnetic field. 
If the chiral asymmetry were comparable in magnitude to the baryon asymmetry of the Universe we would expect $n_{5} \sim 10^{-10} T^3$ or $\mu_{5} \sim 10^{-8} T$.  
For such a small chiral asymmetry, the estimates of $t_{\rm diff}$ in \erefs{eq:tdiff_est}{eq:tdiff_bound} imply that helicity diffusion is too slow to accomplish any significant degree of charge erasure or magnetogenesis.  
Thus, our estimates cannot confirm the assumption of rapid charge erasure via magnetic helicity diffusion, which was utilized in \rref{Long:2013tha} to calculate the production of helical magnetic fields from leptogenesis.  
This is not to say that a primordial magnetic field cannot arise in association with leptogenesis (or baryogenesis) since charge erasure can also be accomplished through the chiral magnetic effect rather than helicity diffusion \cite{Joyce:1997uy,Dvornikov:2011ey,Dvornikov:2012rk,Dvornikov:2013bca,Semikoz:2012ka,Semikoz:2013xkc,Semikoz:2015wsa} (see also \cite{Sabancilar:2013raa,Zadeh:2015oqf,Giovannini:1997eg,Giovannini:1999wv,Giovannini:1999by,Tashiro:2012mf,Anber:2015yca,Boyarsky:2011uy,Boyarsky:2012ex,Boyarsky:2015faa,Fujita:2016igl}).  

Our analysis is sufficiently general that it has applications to Abelian gauge extensions of the Standard Model.  
The additional $\U{1}$ sectors could be related to lepton number, supersymmetric R-charge, or asymmetric dark matter to give a few examples.  
Provided that the $\U{1}$ is unbroken in the early Universe and the associated charge carriers are relativistic, \eref{eq:tdiff_bound} gives the condition for chiral charge erasure by magnetic helicity diffusion. In particular, there could be a wide range of applications for asymmetric dark matter models, where the dark matter could be charged under a gauged dark U(1) and its abundance is determined via non-perturbative processes similar to baryogenesis \cite{Petraki:2013wwa,Zurek:2013wia}.
Finally, although we have focused on the chiral instability of an Abelian plasma, we expect that the energetic argument of \sref{sec:Energetic} will carry over to non-Abelian plasmas as well.  In fact, Refs.~\cite{Akamatsu:2013pjd,Akamatsu:2014yza,Akamatsu:2015kau} recently studied chiral instabilities in a non-Abelian plasma with large chiral asymmetry.  In addition to the well-known sphaleron processes, it might be interesting to study chiral charge erasure through the non-Abelian analog of helicity diffusion.

\begin{acknowledgments}

We would like to especially thank Tanmay Vachaspati for extensive discussions during the early stages of this work.  
We would also like to thank 
Mohamed Anber,
Alexey Boyarsky, 
Yannis Burnier,
Daniel Chung,
Oleg Ruchayskiy, 
Mikhail Shaposhnikov,
and Dam Thanh Son
for stimulating conversations.
AJL is supported at the University of Chicago by the Kavli Institute for Cosmological Physics through grant NSF PHY-1125897 and an endowment from the Kavli Foundation and its founder Fred Kavli.  
ES is supported by Swiss National Science Foundation. 

\end{acknowledgments}


\begin{appendix}
\section{Summary of Green's Functions}\label{app:Greens_func}

First we define the various propagators (see also Chapter 3 of \cite{HuagJauho:2008}).
The time-ordered or causal propagator is
\begin{align}\label{eq:def_Gc}
	i G_{ij}^{\rm c}(x-x^{\prime}) \equiv \langle \mathcal{T} A_i(x) A_j(x^{\prime}) \rangle \per
\end{align}
It reduces to 
\begin{align}
	i G_{ij}^{>}(x-x^{\prime}) & \equiv \langle A_i(x) A_j(x^{\prime}) \rangle \com \label{eq:def_Gg} \\
	i G_{ij}^{<}(x-x^{\prime}) & \equiv \langle A_j(x^{\prime}) A_i(x) \rangle \com \label{eq:def_Gl}
\end{align}
for $(x^0 - x^{\prime 0}) > 0$ and $<0$, respectively.  
The retarded and advanced propagators are 
\begin{align}\label{eq:def_Gr_Ga}
	i G_{ij}^{\rm r}(x-x^{\prime}) & \equiv \Theta(x^0-x^{\prime 0}) \langle [ A_i(x) , A_j(x^{\prime}) ] \rangle \com \\
	i G_{ij}^{\rm a}(x-x^{\prime}) & \equiv -\Theta(x^{\prime 0}-x^0) \langle [ A_i(x) , A_j(x^{\prime}) ] \rangle \per
\end{align}
The symmetrized and anti-symmetrized propagators are 
\begin{align}\label{eq:def_Gs_Ga}
	i G_{ij}^{\rm sym}(x-x^{\prime}) & \equiv \langle \{ A_i(x) , A_j(x^{\prime}) \} \rangle \com \\
	i G_{ij}^{\rm asym}(x-x^{\prime}) & \equiv \langle [ A_i(x) , A_j(x^{\prime}) ] \rangle \per
\end{align}
The various propagators obey the following relations,
\begin{align}\label{eq:G_relations}
	G^{>}_{ij}(x-x^{\prime}) & = G^{<}_{ji}(x^{\prime}-x) \com \nn
	G^{\rm asym}(x-x^{\prime})  & = G^{>}(x-x^{\prime})  - G^{<}(x-x^{\prime})  = G^{\rm r}(x-x^{\prime})  - G^{\rm a}(x-x^{\prime}) \com \nn
	G^{\rm sym}(x-x^{\prime})  & = G^{>}(x-x^{\prime})  + G^{<}(x-x^{\prime})  = \begin{cases} G^{\rm r}(x-x^{\prime})  & \com x^0 > x^{\prime 0} \\ G^{\rm a}(x-x^{\prime})  & \com x^0 < x^{\prime 0} \end{cases} 
	\per
\end{align}
For notational clarity we have suppressed the subscript $ij$ here and below.  

The propagators also have a Fourier space representation,
\begin{align}\label{eq:Gx_to_Gk}
	\wt{G}^{\alpha}(K) = \int \! \ud^4(x-x^{\prime}) \, G^{\alpha}(x-x^{\prime}) \, e^{i K \cdot (x-x^{\prime})}
\end{align}
where $\alpha$ labels the propagator and $K^{\mu} = (\omega,\kvec)$ is the 4-momentum.  
In general $\wt{G}^{\alpha}(K)$ is meromorphic, and we assume that its poles and branch cuts are displaced from the real-$\omega$ axis so that the inverse Fourier transform,
\begin{align}
	G^{\alpha}(x-x^{\prime}) = \int \! \! \frac{\ud^4 k}{(2\pi)^4} \, \wt{G}^{\alpha}(K) \, e^{-i k \cdot (x-x^{\prime})} \com
\end{align}
is well-defined when we integrate $\omega = K^0 \in (-\infty,\infty)$.  
The two-point functions are related to the propagators by 
\begin{subequations}
\begin{align}
	\langle A_{i}(K) A_{j}(K^{\prime})^{\dagger} \rangle & = (2\pi)^4 \delta(K-K^{\prime}) \, i \wt{G}^{>}_{ij}(K) \label{eq:def_Gtildeg} \com \\
	\langle A_{j}(K^{\prime})^{\dagger} A_{i}(K) \rangle & = (2\pi)^4 \delta(K-K^{\prime}) \, i \wt{G}^{<}_{ij}(K) \com \\
	\langle [ A_{i}(K) , A_{j}(K^{\prime})^{\dagger}] \rangle & = (2\pi)^4 \delta(K-K^{\prime}) \, i \wt{G}^{\rm sym}_{ij}(K) \com \\
	\langle \{ A_{i}(K) , A_{j}(K^{\prime})^{\dagger} \} \rangle & = (2\pi)^4 \delta(K-K^{\prime}) \, i \wt{G}^{\rm asym}_{ij}(K) \per
\end{align}
\end{subequations}
Assuming spatial isotropy, the propagators admit a tensor decomposition 
\begin{align}\label{eq:G_tensor_decomp}
	\wt{G}_{ij}^{\alpha}(K) = P_{ij}^{L}(\khat) \, L^{\alpha}(k,\omega) + P_{ij}^{T}(\khat) \, T^{\alpha}(k,\omega) + P_{ij}^{A}(\khat) \, A^{\alpha}(k,\omega) \com
\end{align}
where the longitudinal, transverse, and axial projection operators are defined by 
\begin{align}\label{eq:def_proj_ops}
	P_{ij}^{L}(\khat) = \khat_i \khat_j
	\qquad , \qquad
	P_{ij}^{T}(\khat) = \delta_{ij} - \khat_i \khat_j
	\qquad , \quad \text{and} \qquad
	P_{ij}^{A}(\khat) = -i \epsilon_{ijk} \khat_k
	\com
\end{align}
and $k = \kabs$ and $\khat = \kvec / \kabs$.  
This reduces the propagator to the three functions, $L^{\alpha}$, $T^{\alpha}$, and $A^{\alpha}$.  

Writing the longitudinal and transverse components of the polarization tensor as $\Pi_{l}(k,\omega)$ and $\Pi_{t}(k,\omega)$, the retarded and advanced propagators have a particularly simple pole structure,
\begin{subequations}
\begin{align}
	L^{\rm r,a}(k,\omega) & = 
	\frac{1}{(\omega \pm i \epsilon)^2 - k^2 - \Pi_{l}(k,\omega \pm i\epsilon)} \com \\
	T^{\rm r,a}(k,\omega) & = 
	\frac{1}{(\omega \pm i \epsilon)^2 - k^2 - \Pi_{t}(k,\omega \pm i\epsilon)} \com
\end{align}
\end{subequations}
and $A^{\rm r,a}(k,\omega) = 0$.  
The limit $\epsilon \to 0$ should be taken at the end of the calculation, after the momentum integrals are performed.  
The anti-symmetric propagator is constructed using \eref{eq:G_relations}:  
\begin{subequations}\label{eq:Lasym_Tasym}
\begin{align}
	L^{\rm asym}(k,\omega) & = 
	\frac{1}{(\omega + i \epsilon)^2 - k^2 - \Pi_{l}(k,\omega + i\epsilon)} 
	- \frac{1}{(\omega - i \epsilon)^2 - k^2 - \Pi_{l}(k,\omega - i\epsilon)} \com \\
	T^{\rm asym}(k,\omega) & = 
	\frac{1}{(\omega + i \epsilon)^2 - k^2 - \Pi_{t}(k,\omega + i\epsilon)} 
	- \frac{1}{(\omega - i \epsilon)^2 - k^2 - \Pi_{t}(k,\omega - i\epsilon)} \per
\end{align}
\end{subequations}
If the $\Pi_{l,t}$ are free of branch cuts, then these functions vanish everywhere in the complex-$\omega$ plane as $\epsilon \to 0$.  
If branch cuts are present, then the functions only vanish on the first Riemann sheet.  
These expressions reduce further if $\Pi(k,\omega-i\epsilon) = \Pi(k,\omega+i\epsilon)^{\ast}$, and then 
\begin{subequations}
\begin{align}
	L^{\rm asym}(k,\omega) & = 
	2i \, {\rm Im} \Bigl[ \frac{1}{(\omega + i \epsilon)^2 - k^2 - \Pi_{l}(k,\omega + i\epsilon)} \Bigr] 
	= 2i \, {\rm Im} \bigl[ L^{\rm r}(k,\omega) \bigr] \com \\
	T^{\rm asym}(k,\omega) & = 
	2i \, {\rm Im} \Bigl[ \frac{1}{(\omega + i \epsilon)^2 - k^2 - \Pi_{t}(k,\omega + i\epsilon)} \Bigr] 
	= 2i \, {\rm Im} \bigl[ T^{\rm r}(k,\omega) \bigr] \com
\end{align}
\end{subequations}
where $L^{\rm r}$ and $T^{\rm r}$ are the longitudinal and transverse retarded propagators.

\section{Fluctuation Dissipation Theorem}\label{app:FDT}

The fluctuation-dissipation theorem can be expressed as
\begin{align}\label{eq:FDT_Gg_to_Gasym}
	\wt{G}^{<} & = f_{\BE} \, \wt{G}^{\rm asym} \com
\end{align}
where the Bose-Einstein distribution function is 
\begin{align}\label{eq:def_FBE}
	f_{\BE}(\omega) = \frac{1}{e^{\beta \omega} - 1} = \Bigl( \frac{1}{2} \, \coth \frac{\beta \omega}{2} - \frac{1}{2} \Bigr) \com
\end{align}
and $\beta=1/T$ is the inverse temperature.  
Using the relations in \eref{eq:G_relations}, alternate versions of the fluctuation-dissipation theorem are obtained:
\begin{align}
	\wt{G}^{<} & = \Bigl( \frac{1}{2} \, \coth \frac{\beta \omega}{2} - \frac{1}{2} \Bigr) \wt{G}^{\rm asym} \com\label{eq:FDT_Gl} \\
	\wt{G}^{>} & = \Bigl( \frac{1}{2} \, \coth \frac{\beta \omega}{2} + \frac{1}{2} \Bigr) \wt{G}^{\rm asym} \com\label{eq:FDT_Gg} \\
	\wt{G}^{\rm sym} & = \coth \frac{\beta \omega}{2} \ \wt{G}^{\rm asym} \label{eq:FDT_Gsym}
	\per
\end{align}

For a derivation of the fluctuation-dissipation theorem see chapter 3 of \cite{HuagJauho:2008} (also chapter 1 of \cite{Sitenko:1967} or chapter 11 of \cite{Akhiezer:1975a,Akhiezer:1975b}).  
One can obtain \eref{eq:FDT_Gg_to_Gasym} in the following way.  
sThe thermally averaged unequal time two point functions, \erefs{eq:def_Gg}{eq:def_Gl}, can be written as 
\ba
iG_{ij}^{>}(x-x') &=& \frac{1}{Z} \sum_{n} \bra{n} e^{-\beta H} A_i(x) A_j(x')  \ket{n} \com\\
iG_{ij}^{<}(x-x') &=& \frac{1}{Z} \sum_{n} \bra{n} e^{-\beta H} A_j(x') A_i(x)  \ket{n} \com
\ea
where $Z ={\rm Tr}~e^{-\beta H}$ is the partition function, $H$ is the Hamiltonian of the system, $H \ket{n} = E_n \ket{n}$ and $\left\{ \ket{n} \right\}$ form an orthonormal basis of the energy eigenstates. 
The Heisenberg picture field operators can be written as $A\xt = e^{iH t} A({\bm x},0) e^{-iHt}$, and we will drop the spatial index for notational simplicity.  
Inserting an identity $1 = \sum_m \ket{m} \bra{m}$ leads to 
\ba
iG^{>}(x-x') &=& \frac{1}{Z} \sum_{m,n} e^{-\beta E_n}~ e^{i (E_n-E_m)(t-t')} \mathcal{A}_{nm}({\bm x})~ \mathcal{A}_{mn}({\bm x}') \com \label{G>t}\\
iG^{<}(x-x') &=& \frac{1}{Z} \sum_{m,n} e^{-\beta E_n} ~e^{i (E_n-E_m)(t-t')} \mathcal{A}_{nm}({\bm x}) ~\mathcal{A}_{mn}({\bm x}') ~e^{-\beta (E_m-E_n)} \com \label{G<t}
\ea
where we defined $\mathcal{A}_{nm}({\bm x}) \equiv \bra{n} A({\bm x},0) \ket{m}$, and we have also replaced $m \to n$ to get the expression for $G^{<}(x-x')$ in the second line.  
Note that the expressions under the sum only differ by a factor of $e^{-\beta (E_m-E_n)}$.  
Upon taking the temporal Fourier transforms of (\ref{G>t}) and (\ref{G<t}), we obtain 
\ba
i\wt{G}^{>} ({\bm x}-{\bm x}',\omega) &=& \frac{2\pi}{Z} \sum_{m,n} e^{-\beta E_n}~ \mathcal{A}_{nm}({\bm x})~ \mathcal{A}_{mn}({\bm x}')~\delta(\omega+ E_n-E_m) \com \label{G>w}\\  
i\wt{G}^{<} ({\bm x}-{\bm x}',\omega) &=& \frac{2\pi}{Z} \sum_{m,n} e^{-\beta E_n}~ \mathcal{A}_{nm}({\bm x})~ \mathcal{A}_{mn}({\bm x}')~\delta(\omega+ E_n-E_m)~e^{-\beta (E_m-E_n)} \label{G<w} \per
\ea
The Dirac delta function $\delta(\omega+ E_n-E_m)$ in the above expressions guarantees that $E_m -E_n = \omega$, hence, (\ref{G<w}) can also be written in terms of (\ref{G>w}) as
\be \label{FDT}
\wt{G}^{<}({\bm x}-{\bm x}',\omega)  = e^{-\beta \omega} \wt{G}^{>}({\bm x}-{\bm x}',\omega) \per
\ee
We finally obtain \eref{eq:FDT_Gg_to_Gasym} by taking the spatial Fourier transform and using the relations in \eref{eq:G_relations}.  

\section{Thermal Fluctuations in a Relativistic Plasma}\label{app:Rel_Plasma}

For a relativistic plasma in the high-temperature (hard thermal loop) approximation $m, \omega,k \ll T$, the longitudinal and transverse components of the photon polarization tensor are \cite{Weldon:1982aq}
\begin{align}
	\Pi_{l}(k,\omega) & 
	= \frac{1}{a^2} \left(1 - \frac{\omega^2}{k^2} \right) \left[ 1 + \frac{1}{2} \frac{\omega}{k} \ln \frac{\omega - k}{\omega + k} \right] \com
	\label{eq:Pil_highT}
	\\
	\Pi_{t}(k,\omega) & 
	= \frac{1}{2a^2} \frac{\omega^2}{k^2} \left[ 1 - \frac{1}{2} \frac{k}{\omega} \left( 1 - \frac{\omega^2}{k^2} \right) \ln \frac{\omega - k}{\omega + k} \right] 
	\label{eq:Pit_highT}
	\per
\end{align}
The parameter $a$ is the Debye length, and the logarithm is a Legendre function of the second kind $Q_0(z) = (1/2) \log ( 1 + z ) / (1-z)$ with $z = k / \omega$.  
Both $\Pi_l$ and $\Pi_t$ have a branch cut along $\omega \in (-k,k)$.  

\begin{figure}[t]
\begin{center}
\includegraphics[width=0.48\textwidth]{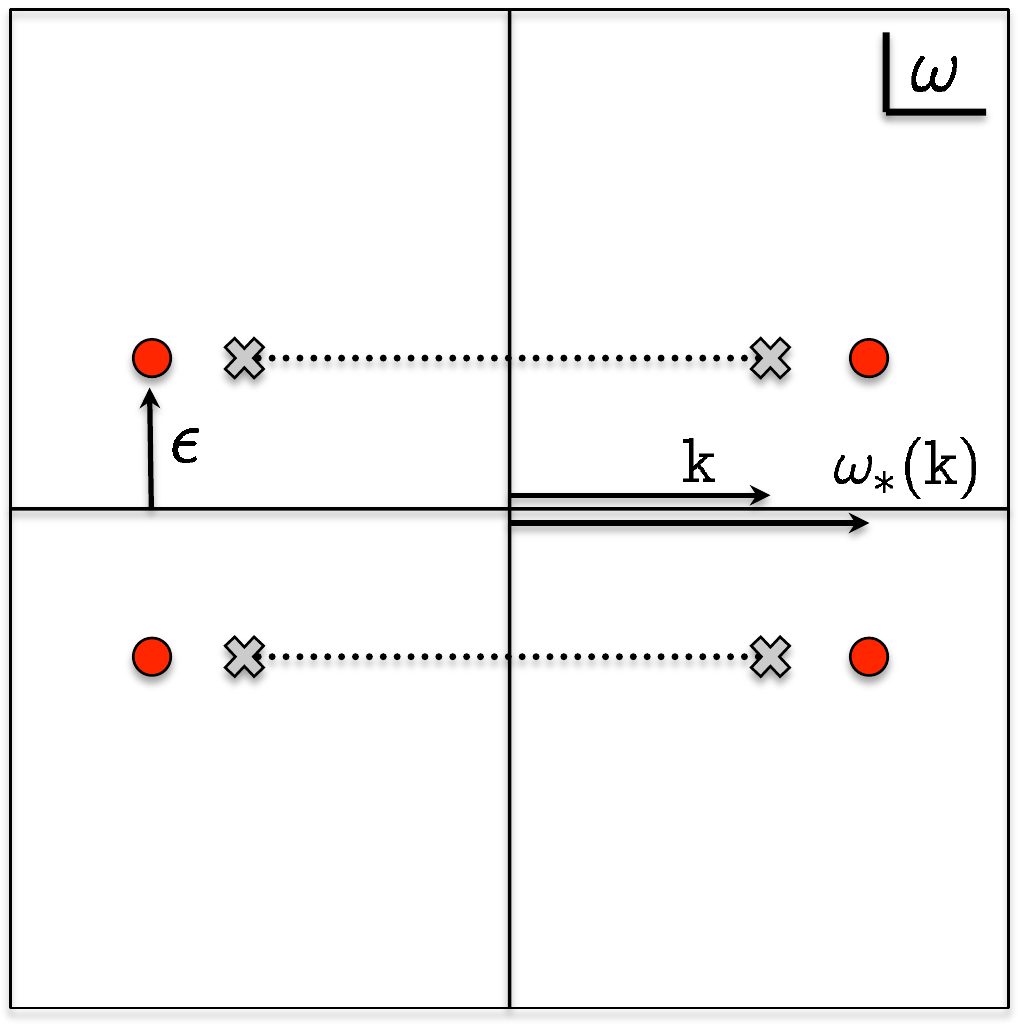} 
\caption{
\label{fig:complex_plane}
The analytic structure of $L^{\rm asym}(k,\omega)$ and $T^{\rm asym}(k,\omega)$ in the complex $\omega$-plane.  The red dots indicate first-order poles, the gray crosses indicate branch points, and the dotted lines indicate branch cuts.  
}
\end{center}
\end{figure}

Using \erefs{eq:Pil_highT}{eq:Pit_highT} we construct the antisymmetric propagators $L^{\rm asym}$ and $T^{\rm asym}$ from \eref{eq:Lasym_Tasym}.  
Both functions have the same analytic structure, which is shown in \fref{fig:complex_plane}.  
The poles and cut in the lower (upper) half plane arise from the first (second) terms in \eref{eq:Lasym_Tasym}, corresponding to the retarded (advanced) propagator.  
As we take $\epsilon \to 0$ the functions vanish everywhere on the first Riemann sheet apart from the region pinched off between the branch cuts.  
This leaves \cite{Sitenko:1967} 
\begin{align}
	i L^{\rm asym}(k,\omega) & = 
	\frac{-2 \, {\rm I}_l(k,\omega)}{\bigl[ \omega^2 - k^2 - {\rm R}_l(k,\omega) \bigr]^2 + \bigl[ {\rm I}_l(k,\omega) \bigr]^2 } 
	\label{eq:Lasym} \com \\
	i T^{\rm asym}(k,\omega) & = 
	\frac{-2 \, {\rm I}_t(k,\omega)}{\bigl[ \omega^2 - k^2 - {\rm R}_t(k,\omega) \bigr]^2 + \bigl[ {\rm I}_t(k,\omega) \bigr]^2 } 
	\label{eq:Tasym} \com
\end{align}
where
\begin{subequations}
\begin{align}
	{\rm R}_l(k,\omega) & = \frac{1}{a^2} \left(1 - \frac{\omega^2}{k^2} \right) \left[ 1 + \frac{1}{2} \frac{\omega}{k} \ln \frac{|\omega - k|}{|\omega + k|} \right] \com \\
	{\rm I}_l(k,\omega) & = \frac{1}{a^2} \left(1 - \frac{\omega^2}{k^2} \right) \frac{1}{2} \frac{\omega}{k} \pi \, \Theta \bigl[ (k-\omega) (\omega+k) \bigr] \com \\
	{\rm R}_t(k,\omega) & = \frac{1}{2a^2} \frac{\omega^2}{k^2} \left[ 1 - \frac{1}{2} \frac{k}{\omega} \left( 1 - \frac{\omega^2}{k^2} \right) \ln \frac{|\omega - k|}{|\omega + k|} \right] \com \\
	{\rm I}_t(k,\omega) & = - \frac{1}{2a^2} \frac{\omega^2}{k^2} \frac{1}{2} \frac{k}{\omega} \left( 1 - \frac{\omega^2}{k^2} \right) \pi \, \Theta \bigl[ (k-\omega) (\omega+k) \bigr]  \per
\end{align}
\end{subequations}
Here $\Theta(x)$ is the Heaviside step function, which equals $1$ for $x>0$ and $0$ otherwise.  
These functions are nothing more than the real and imaginary parts of the $\Pi_{l,t}$ evaluated along the real axis and above the branch cut from $-k < \omega < k$.  
We can also write
\begin{align}\label{eq:Lasym_Tasym_rel}
	L^{\rm asym}(k,\omega) & = 
	2i \, {\rm Im} \left[ \frac{1}{\omega^2 - {\bf k}^2 - {\rm R}_l(k,\omega) - i {\rm I}_l(k,\omega) } \right] \com \\
	T^{\rm asym}(k,\omega) & = 
	2i \, {\rm Im} \left[ \frac{1}{\omega^2 - {\bf k}^2 - {\rm R}_t(k,\omega) - i {\rm I}_t(k,\omega) } \right]
\end{align}
for $\omega$ along the real axis where the ${\rm R}$'s and ${\rm I}$'s are real.  
Finally the fluctuation-dissipation theorem in \eref{eq:FDT_Gg} gives
\begin{align}\label{eq:Lasym_Tasym_rel}
	L^{>}(k,\omega) & = 
	2i \Bigl( \frac{1}{\beta \omega} + \frac{1}{2} \Bigr) \, {\rm Im} \left[ \frac{1}{\omega^2 - {\bf k}^2 - {\rm R}_l(k,\omega) - i {\rm I}_l(k,\omega) } \right] \com \\
	T^{>}(k,\omega) & = 
	2i \Bigl( \frac{1}{\beta \omega} + \frac{1}{2} \Bigr) \, {\rm Im} \left[ \frac{1}{\omega^2 - {\bf k}^2 - {\rm R}_t(k,\omega) - i {\rm I}_t(k,\omega) } \right] \com
\end{align}
where we have used $\beta \omega \ll 1$ and expanded $\coth \beta \omega/2$.  

The root-mean-square (rms) amplitude of transverse gauge field fluctuations is estimated as 
\begin{align}
	A_{\rm rms}(k,\omega) 
	& \sim \lim_{\omega^{\prime} \to \omega} \lim_{{\bf k}^{\prime} \to {\bf k}} \sqrt{ \frac{\omega}{2\pi} \frac{k^3}{(2\pi)^3} \frac{\omega^{\prime}}{2\pi} \frac{k^{\prime 3}}{(2\pi)^3} P_{ij}^{T}(\khat) \langle A_i(K) A_j(K^{\prime})^{\dagger} \rangle } \com
\end{align}
where the transverse projection operator $P_{ij}^{T}(\khat)$ was defined in \eref{eq:def_proj_ops}.  
The two-point function is simply $(2\pi)^4 \delta(K-K^{\prime}) i\wt{G}_{ii}^{>}(K)$ from \eref{eq:def_Gtildeg}.  
Using the fluctuation-dissipation theorem, \eref{eq:FDT_Gg}, we write $\wt{G}^{>}$ in terms of $\wt{G}^{\rm asym}$ to obtain 
\begin{align}
	A_{\rm rms}(k,\omega) 
	& \sim\sqrt{ \frac{\omega}{2\pi} \frac{k^3}{(2\pi)^3} \left( \frac{1}{2} \coth \frac{\beta \omega}{2} + \frac{1}{2} \right) P_{ij}^{T}(\khat) \, i \wt{G}_{ij}^{\rm asym}(K) } \per
\end{align}
In the high-temperature limit, $\beta \omega \ll 1$, the parenthetical factor is simply $1/\beta \omega$.  
Using now the tensor decomposition in \eref{eq:G_tensor_decomp} and the expression for $T^{\rm asym}$ in \eref{eq:Tasym} gives 
\begin{align}\label{eq:Arms_appendix}
	A_{\rm rms}(k,\omega) 
	& \sim \sqrt{ \frac{\omega}{2\pi} \frac{k^3}{(2\pi)^3} \frac{1}{\beta \omega} \, i T^{\rm asym}(k,\omega) } \\
	& = \sqrt{\frac{T}{a}} \frac{(ak)^{3/2}}{\sqrt{2} \pi ^{3/2}}\sqrt{\frac{x \left(1-x^2\right)}{\pi ^2 x^2 \left(1-x^2\right)^2+\left[2 x^2+4 a^2k^2 \left(1-x^2\right)+x \left(1-x^2\right) \ln \frac{1+x}{1-x}\right]^2}}
	\nonumber
\end{align}
where $x \equiv \omega / k$ and we assume $0<x<1$.  
For $|\omega| \ll k$ and $ak \ll 1$ we can expand this expression to obtain
\begin{align}\label{eq:Arms_highT_app}
	A_{\rm rms}(k,\omega) \sim 
	\sqrt{\frac{T}{a}} \left( 
	\frac{1}{4\pi k} \sqrt{ \frac{1}{2 \pi a} } \omega^{1/2} 
	- \frac{\pi}{128 a^4 k^7} \sqrt{ \frac{1}{2 \pi a} } \omega^{5/2} 
	+ \cdots 
	\right) \per
\end{align}

\end{appendix}

\bibliographystyle{JHEP}
\bibliography{helicity_fluct}

\end{document}